\documentclass[11pt]{article}

\usepackage{amsmath,amssymb}

\makeatletter

 \@addtoreset{equation}{section}
\makeatother

\newcommand{\Pd}[2]{\frac{\partial{#1}}{\partial{#2}}}
\newcommand{\ltd}[1]{\frac{d}{d{#1}}}

\newcommand{\del}{\partial}
\newcommand{\half}{\frac{1}{2}}
\newcommand{\LS}{\ \ \ \ \ \ \ \ \ \ }
\newcommand{\ls}{\ \ \ \ \ }
\newcommand{\ve}{\varepsilon}
\newcommand{\ol}{\overline}
\newcommand{\ul}{\underline}
\newcommand{\dps}{\displaystyle}
\newcommand{\kahler}{K\"{a}hler }

\newcommand{\bsubeq}{\begin{subequations}}
\newcommand{\esubeq}{\end{subequations}}
\newcommand{\vs}[1]{\vspace{#1 mm}}

\newcommand{\tr}{{\rm tr}}

\begin{document}

\allowdisplaybreaks{

\setcounter{page}{0}

\begin{titlepage}

{\normalsize
\begin{flushright}
OU-HET 395 \\
PURD-TH-01-04 \\
{\tt hep-th/0110216} \\
October 2001
\end{flushright}
}

\vs{2}

\begin{center}
{\Huge Gauge Theoretical Construction 

of Non-compact Calabi-Yau Manifolds}

\vs{7}

\bigskip
{\renewcommand{\thefootnote}{\fnsymbol{footnote}}
{\Large\bf Kiyoshi Higashijima$^1$\footnote{
     E-mail: {\tt higashij@phys.sci.osaka-u.ac.jp}},
 Tetsuji Kimura$^1$\footnote{
     E-mail: {\tt t-kimura@het.phys.sci.osaka-u.ac.jp}}
 and Muneto Nitta$^2$\footnote{
     E-mail: {\tt nitta@physics.purdue.edu}}
}}

\vs{2}

{\large\sl
$^1$ Department of Physics,
Graduate School of Science, Osaka University, \\
Toyonaka, Osaka 560-0043, Japan \\

$^2$ Department of Physics, Purdue University, West Lafayette, IN
47907-1396, USA
}

\end{center}


\vs{7}

\begin{abstract}

We construct 
the non-compact Calabi-Yau manifolds
interpreted as the complex line bundles over the Hermitian symmetric
spaces. 
These manifolds are the various generalizations of 
the complex line bundle over ${\bf C}P^{N-1}$.
Imposing an F-term constraint on 
the line bundle over ${\bf C}P^{N-1}$, 
we obtain the line
bundle over the complex quadric surface $Q^{N-2}$.
On the other hand,
when we promote the $U(1)$ gauge symmetry in ${\bf C}P^{N-1}$ 
to the non-abelian gauge group $U(M)$, 
the line bundle over the Grassmann manifold 
is obtained.
We construct the non-compact Calabi-Yau manifolds 
with isometries of exceptional groups, 
which we have not discussed in 
the previous papers.
Each of these manifolds contains the resolution parameter which
controls the size of the base manifold,
and the conical singularity appears 
when the parameter vanishes.

\end{abstract}

\end{titlepage}


\section{Introduction} \label{intro}

Two-dimensional ${\cal N}=2$ supersymmetric nonlinear sigma models 
on Calabi-Yau (Ricci-flat K\"{a}hler) manifolds 
are considered as the most important models 
of the superstring theory~\cite{Zu,finite,GVZ,NS}. 
While the supergravity is 
now regarded as the low-energy effective theory of string/M-theory, 
the supergravity solutions preserving some of the supersymmetry 
are considerable issues in the exploration both
of perturbative and of non-perturbative properties 
in string and M-theory. 
In particular,
the AdS/CFT correspondence is one of the most powerful conjectures
to obtain these properties.
A useful example of this correspondence has been to study D3-branes on
the manifold with or without conical singularities.  
Recently, 
some resolution procedures of singularities have been extensively
discussed 
(see, for example, \cite{KleTsey,KleStr,PT,CLP,PapTsey,CGLP} 
and references therein),
because 
such non-singular examples may provide important supergravity dual
solutions of four-dimensional ${\cal N}=1$ super-Yang-Mills theory in
the infrared regime.  
It is also important to understand real manifolds 
with and without singularities 
in recent study of compactification of M-theory on 
manifolds with
$G_2$ and $Spin(7)$ holonomies, which presents 
field theories with less supersymmetry 
(see, for example,
\cite{Spin7,Ach,ES,AW,GLPS}). 
Returning to
the nonlinear sigma models and the string world-sheet 
on Calabi-Yau manifolds, 
it is important that we represent these manifolds in 
the complex coordinates from requirement of 
manifestation of supersymmetry. 

In \cite{HKN1},
we presented a simple construction 
of the $O(N)$ symmetric Ricci-flat \kahler metric 
which coincides with the Stenzel metric 
on the cotangent bundle over $S^{N-1}$~\cite{St,CGLP}.
The conical singularity of the conifold is resolved 
by $S^{N-1}$ with a radius being the deformation parameter.
Low dimensional manifolds coincide with 
the Eguchi-Hanson gravitational instanton~\cite{EH}  
or the six-dimensional deformed conifold~\cite{conifold,conifold2}.
A new way of replacing the node of the conifold, 
which is 
different from neither the small resolution 
nor the deformation discussed in 
\cite{conifold,conifold2}, 
was found in \cite{PT,HKN2}.
The new Ricci-flat \kahler manifold is identified 
as the complex line bundle over 
the complex quadric surface 
$Q^{N-2} = SO(N)/[SO(N-2) \times U(1)]$~\cite{HKN2}. 
In \cite{HKN3},
we constructed new manifolds 
whose conical singularities are resolved 
by other Hermitian symmetric spaces (HSS, see \cite{HN1,HN2})
with classical groups, and found that 
these manifolds are the complex line bundles over 
the Grassmann manifold 
$G_{N,M}=SU(N)/[SU(N-M)\times U(M)]$, 
$SO(2N)/U(N)$ and $Sp(N)/U(N)$. 
All of these are gerenalizations of 
the line bundle over 
the complex projective space ${\bf C}P^{N-1}
= SU(N)/[SU(N-1)\times U(1)]$~\cite{Ca}. 
In this paper
we present the new conifolds with isometries of 
$E_6$ and $E_7$.   
The conical singularities of these conifolds 
are resolved by the HSS of exceptional groups,  
$E_6/[SO(10) \times U(1)]$ and
$E_7/[E_6 \times U(1)]$, 
and new manifolds are identified 
as the complex line bundles over 
these HSS.
Then we summarize the non-compact Calabi-Yau manifolds
interpreted as the complex line bundles over all of the HSS.

This paper is organized as follows.
In section \ref{Calabi-section}
we discuss the ${\bf C}P^{N-1}$ model and 
the complex line bundle over ${\bf C}P^{N-1}$. 
In section \ref{LQ},
we discuss the conifold as 
the complex line bundle 
over the complex quadric surface $Q^{N-2}$, 
which is the simple generalization of 
the line bundle over ${\bf C}P^{N-1}$.
In section \ref{LE},
we discuss the {\it new} conifolds 
with isometries of the exceptional groups, 
$E_6$ and $E_7$.
As in the conifold, the conical singularity of 
$E_6$ or $E_7$ conifold 
is resolved by the HSS of the exceptional group, 
$E_6/[SO(10)\times U(1)]$ or $E_7/[E_6\times U(1)]$. 
In section \ref{LG},
we review the gauge theoretical construction of 
the line bundle over the Grassmann manifold $G_{N,M}$ 
and its generalizations;
the line bundles over $SO(2N)/U(N)$ and $Sp(N)/U(N)$.
These models are 
the non-abelian generalizations of 
the line bundle over ${\bf C}P^{N-1}$. 
Section \ref{conclusion} is devoted to the conclusion.
In appendix \ref{SO10},
we define the $SO(10)$ $\gamma$-matrices 
and the charge conjugation matrix 
in the Weyl spinor basis.
These representations are useful in performing explicit calculations 
in section \ref{LE}.
In appendix \ref{e-algebra},
we review the $E_6$ and $E_7$ algebras.


\section{Complex Line Bundle over ${\bf C}P^{N-1}$}
\label{Calabi-section}

In this section
we review the gauge theoretical construction 
of the ${\bf C}P^{N-1}$ model~\cite{DDL}.
Then we derive 
the complex line bundle over ${\bf C}P^{N-1}$~\cite{Ca}.

\ul{\sl ${\bf C}P^{N-1}$ Model}. \hfil\break
We consider 
the global symmetry $G = SU(N)$ and the $U(1)_{\rm local}$ symmetry. 
We introduce chiral superfields  
$\vec{\phi}(x, \theta, \bar{\theta})$,  
belonging to the fundamental representation of $SU(N)$, 
and a vector superfield $V (x, \theta, \bar{\theta})$ 
of the $U(1)_{\rm local}$ gauge symmetry 
as an auxiliary field.
The gauge transformation of $U(1)_{\rm local}$ is given by
\begin{align}
\vec{\phi} \ &\to \ 
\vec{\phi}' 
\ = \ 
\vec{\phi} e^{-i \Lambda} \; , \ \ \ 
e^V \ \to \ e^{V'} 
\ = \ 
e^{i \Lambda} e^V e^{-i \Lambda^{\dagger}} \; , \label{gauge-CPN}
\end{align}
where $\Lambda (x, \theta, \bar{\theta})$ is a chiral superfield.
Note that the local invariance group is enlarged 
to  the complexification $U(1)_{\rm local}^{\bf C}$ 
of the $U(1)_{\rm local}$ gauge group,
because the scalar component of 
$\Lambda (x, \theta, \bar{\theta})$ is
a complex field.
The Lagrangian invariant under the global $SU(N)$ and 
the $U(1)_{\rm local}$
symmetries is given by
\begin{align}
{\cal L} 
\ &= \ 
\int \! d^4 \theta 
 (\vec{\phi}^{\dagger} \vec{\phi} e^V - c V)  
 \; . \label{CPN}
\end{align}
The integrand is a \kahler potential, 
in which $c$ is a real positive constant, 
and the term $c V$ is called a Fayet-Iliopoulos (FI) 
D-term.
Integrating out the vector superfield $V$, 
we obtain the following \kahler potential: 
\begin{align}
 \Psi (\vec{\phi}, \vec{\phi}^{\dagger}) 
\ &= \ 
 c \log \big( \vec{\phi}^{\dagger} \vec{\phi} \big) \; , 
\label{CPN-kahler-1}
\end{align}
where we have omitted constant terms because 
they disappear under the integration over $\theta$. 
The complexified gauge symmetry (\ref{gauge-CPN}) can be 
fixed by choosing 
\begin{align}
\vec{\phi} \ &= \ \begin{pmatrix}
 1 \cr 
 \varphi^i 
\end{pmatrix} \; , \label{csf-CPN}
\end{align}
where $\varphi^i (x, \theta, \bar{\theta})$ are chiral superfields ($i
= 1, 2, \cdots, N-1$).
Substituting this into (\ref{CPN-kahler-1}),
we obtain the \kahler potential for 
the Fubini-Study metric on ${\bf C}P^{N-1}$:
\begin{align}
 \Psi(\varphi, \varphi^{\dagger}) 
 \ &= \ 
 c \log \big( 1 + |\varphi^i|^2 \big) \; , \label{CPN-Kahler}
\end{align}
where the summation over the index $i$ is implied. 
Our choice of gauge (\ref{csf-CPN}) breaks global $SU(N)$ symmetry down
to $SU(N-1) \times U(1)$ 
which preserves the vacuum expectation value 
$\langle \phi^1 \rangle = 1$.
$\varphi^i$ represent the massless Nambu-Goldstone bosons
corresponding to the coordinates of 
the coset manifold $G/H = SU(N)/[SU(N-1) \times U(1)]$.

\ul{\sl Complex Line Bundle over ${\bf C}P^{N-1}$}. \hfil\break
Now let us construct the non-compact Calabi-Yau manifold. 
In general,
the Ricci-flat condition is a set of partial differential equations 
difficult to solve.
If we impose a global symmetry, however,
this condition often reduces to a more tractable ordinary differential
equation.
In our example,
we assume the \kahler potential ${\cal K}$ is a function of a single
variable
\begin{align}
X (\vec{\phi}, \vec{\phi}^{\dagger}) 
\ &\equiv \ 
\log \vec{\phi}^{\dagger} \vec{\phi} \; , \label{inv-X-Calabi}
\end{align}
which is invariant under the global $SU(N)$ symmetry.
Here the logarithm in the definition of $X$ is just a convention.

To construct the line bundle over ${\bf C}P^{N-1}$, 
we treat $U(1)_{\rm local}$ symmetry in ${\bf C}P^{N-1}$ 
as a global symmetry
by removing the vector superfield $V$.
Although we do not have $U(1)_{\rm local}$ symmetry,
it is convenient to use the following parametrization 
to compare with the ${\bf C}P^{N-1}$ model,
\begin{align}
\vec{\phi} \ &= \ \sigma \begin{pmatrix}
 1 \cr 
 \varphi^i 
\end{pmatrix} \; , \label{non-cpt-csf}
\end{align}
where $\varphi^i (x , \theta, \bar{\theta})$ and 
$\sigma (x, \theta , \bar{\theta})$ are chiral superfields.
The invariant superfield $X$ is decomposed as
\begin{align}
X \ &= \ 
\log |\sigma|^2 + \Psi \; , \ \ \ 
\Psi \ = \ \log \xi \; , \ \ \ 
\xi 
\ \equiv \ 
1 + |\varphi^i|^2 \; . \label{inv-field}
\end{align}
$\Psi$ is the \kahler potential (\ref{CPN-Kahler}) 
of ${\bf C}P^{N-1}$. 
(Hereafter we set $c=1$.) 

We make a comment on the symmetry breaking.
We find that 
the total space can be regarded as 
\begin{align}
{\bf R} \times \frac{SU(N)}{SU(N-1)} \; ,
\end{align}
at least locally.
The part of $SU(N)/SU(N-1)$ is parametrized by the Nambu-Goldstone bosons
arising from the spontaneous breaking of the global symmetry $SU(N)$ down
to $SU(N-1)$,
whereas the factor ${\bf R}$ is parametrized by the so-called
quasi-Nambu-Goldstone boson~\cite{HN1,NLR,Ni1}.

Let us calculate the \kahler metric and the Ricci tensor. From 
now on we use the same {\it letters} for chiral superfields 
{\it and} their complex scalar components.
The metric of the complex coordinates is defined by $g_{\mu
\nu^*} = \del_{\mu} \del_{\nu^*} {\cal K}$, 
where we express
holomorphic coordinates by $z^{\mu} = (\sigma, \varphi^i)$ 
and the differentiation with respect to them  
by $\del_{\mu} = \del / \del z^{\mu}$.
The explicit expression of the metric is
\bsubeq\label{metric}
\begin{align}
g_{\mu \nu^*} \ &= \ \left(
\begin{array}{cc}
g_{\sigma \sigma^*} & g_{\sigma j^*} \\
g_{i \sigma^*} & g_{i j^*}
\end{array} \right) \; ,
\end{align}
with each block being
\begin{align}
g_{\sigma \sigma^*} 
\ &= \ 
{\cal K}'' \Pd{X}{\sigma} \Pd{X}{\sigma^*} \; , \ \ \ 
g_{\sigma j^*} 
\ = \ 
{\cal K}'' \Pd{X}{\sigma} \Pd{X}{\varphi^{*j}} \; , \ \ \
g_{i j^*} 
\ = \ 
{\cal K}'' \Pd{X}{\varphi^i} \Pd{X}{\varphi^{*j}} 
+ {\cal K}' \frac{\del^2 X}{\del \varphi^i \del \varphi^{*j}} \; , 
\end{align}
\esubeq
where the prime denotes the differentiation with respect to
the argument $X$. 
Here we have used the following equations: 
$\frac{\del^2 X}{\del \sigma \del \sigma^*} =
\frac{\del^2 X}{\del \sigma \del \varphi^{*j}} = 0$ ($\sigma \neq 0$).
The determinant of this metric is given by 
\begin{align}
\det g_{\mu \nu^*} 
\ &= \ 
\frac{1}{|\sigma|^2} {\cal K}'' \cdot \det_{i,j} 
\Big( 
{\cal K}' \frac{\del^2 X}{\del \varphi^i \del \varphi^{*j}} 
\Big) \; . \label{metric-determinant}
\end{align}
Since the Ricci tensor is defined 
by $(Ric)_{\mu \nu^*} = - \del_{\mu}
\del_{\nu^*} \log \det g_{\kappa \lambda^*}$, 
the Ricci-flat condition $(Ric)_{\mu \nu^*} = 0$ 
implies
\begin{align}
\det g_{\mu \nu^*} \ &= \ \mbox{(constant)} \times |F|^2 \; ,
\label{RF-condition}
\end{align}
where $F$ is a holomorphic function.
In order to obtain a concrete expression of 
the determinant (\ref{metric-determinant}),
we need some derivatives of $X$ with respect to $\varphi^i$: 
\begin{align}
\Pd{X}{\varphi^{*j}} 
\ &= \ 
\del_{j^*} \Psi \ = \ \varphi^j \xi^{-1} \; , \ \ \ 
\frac{\del^2 X}{\del \varphi^i \del \varphi^{*j}} 
\ = \ 
\del_i \del_{j^*} \Psi 
\ = \ 
\delta_{i j} \xi^{-1} - \varphi^{* i} \varphi^{j} \xi^{-2} \; . 
\label{deriv-Calabi}
\end{align}
Note that the second quantity is just 
the Fubini-Study metric of ${\bf C}P^{N-1}$. 

Although
it is not difficult to evaluate the determinant
(\ref{metric-determinant}),
it is far simpler if we use its symmetry property.
The determinant is invariant under the transformation $\varphi \to g
\varphi$, 
where $g$ belongs to a complex isotropy group 
$SU(N-1)^{\bf C} = SL(N-1, {\bf C})$ 
that leaves vacuum expectation values invariant.
With suitable choice of $g$, 
we can assume only $\varphi^1$ has a non-vanishing value 
as a vacuum expectation value.
Then the second derivative of $X$, (\ref{deriv-Calabi}), 
reduces to an $(N-1) \times (N-1)$
diagonal matrix:
\begin{align}
\frac{\del^2 X}{\del \varphi^i \del \varphi^{*j}} 
\ &= \ 
{\rm diag.} 
\big( 
\xi^{-2} , \underbrace{\xi^{-1}, \xi^{-1} , \cdots , \xi^{-1}}_{N-2} 
\big) \; . 
\end{align}
Substituting this into (\ref{metric-determinant}),
we have
\begin{align}
\det g_{\mu \nu^*} 
\ &= \
|\sigma|^{2N - 2} e^{-N X} {\cal K}'' ({\cal K}')^{N-1} \; ,
\end{align}
where we have used the relation $\xi = |\sigma|^{-2} e^X$ from
(\ref{inv-field}).

The Ricci-flat condition (\ref{RF-condition})  
reduces to the desired ordinary differential equation 
\begin{align}
e^{- N X} \ltd{X} ({\cal K}')^N \ &= \ a \; , \label{RF-Calabi}
\end{align}
where $a$ is a constant. 
The solution of (\ref{RF-Calabi}) for ${\cal K}'$ is 
\begin{align}
{\cal K}' \ &= \ \big( \lambda e^{N X} + b \big)^{\frac{1}{N}} \; ,
\label{sol-RF-Calabi}
\end{align}
where $\lambda$ is a constant related to $a$ and $N$, and 
$b$ is an integration constant interpreted as a {\it resolution
parameter} of the conical singularity.
Although it is sufficient for us to obtain the metric from
(\ref{sol-RF-Calabi}), 
we can calculate the \kahler potential itself:
\begin{align}
{\cal K} (X) 
\ &= \ 
\big( \lambda e^{NX} + b \big)^{\frac{1}{N}} 
+ b^{\frac{1}{N}} 
\cdot I \big( b^{- \frac{1}{N}} 
\big( \lambda e^{N X} + b \big)^{\frac{1}{N}} ; N \big) \; , 
\label{kahler-Calabi}
\end{align}
where the function $I(y;n)$ is defined by
\begin{align}
I (y; n) 
\ \equiv \ 
  \int^{y} \! \frac{dt}{t^n - 1} 
\ &= \ \frac{1}{n} \Big[ \log \big( y - 1 \big) 
    - \frac{1 + (-1)^n}{2} 
    \log \big( y + 1 \big) \Big] \nonumber \\
& \ \ \ \ 
 + \frac{1}{n} \sum_{r=1}^{[\frac{n-1}{2}]} \cos \frac{2 r \pi}{n} 
\cdot \log \Big( y{}^2 - 2 y \cos \frac{2 r \pi}{n} + 1 \Big) 
\nonumber \\
\ & \ \ \ \ + \frac{2}{n} \sum_{r=1}^{[\frac{n-1}{2}]} 
\sin \frac{2 r \pi}{n} 
\cdot \arctan \Big[ \frac{\cos (2 r \pi / n) - y}
                    {\sin (2 r \pi /n) } \Big] \; . \label{fn-I}
\end{align}

Let us calculate the components of the metric tensor. 
The component $g_{\sigma \sigma^*}$ is calculated as
\begin{align}
g_{\sigma \sigma^*} 
\ &= \ 
\lambda \big( \lambda e^{N X} + b \big)^{\frac{1-N}{N}} 
e^{N \Psi} |\sigma|^{2 N - 2} \; ,
\end{align}
where $\Psi$ is the \kahler potential obtained 
in (\ref{CPN-Kahler}) or (\ref{inv-field}).
This metric has a singularity at the $\sigma = 0$ surface: $g_{\sigma
\sigma^*} = 0$.
However this singularity is just a coordinate singularity of $z^{\mu}
= (\sigma, \varphi^i)$.
To find a regular coordinate system we perform the coordinate
transformation 
\begin{align}
\rho \ &\equiv \ \sigma^N / N \; , \label{rho-Calabi}
\end{align}
with $\varphi^i$ being unchanged. 
Then the metric in the regular coordinates $z'{}^{\mu} = (\rho,
\varphi^i)$ is calculated as
\begin{align}
g_{\rho \rho^*} 
\ &= \ 
\lambda \big( \lambda e^{N X} + b \big)^{\frac{1-N}{N}} e^{N \Psi} 
\; , \nonumber \\
g_{\rho j^*} 
\ &= \ 
\lambda N \big( \lambda e^{N X} + b \big)^{\frac{1-N}{N}} 
e^{N \Psi} \rho^* \cdot \del_{j^*} \Psi \; ,
\label{regular-Calabi} \\
g_{i j^*} 
\ &= \ 
\lambda N^2 \big( \lambda e^{N X} + b \big)^{\frac{1-N}{N}} 
e^{N \Psi} |\rho|^2 \cdot \del_i \Psi \del_{j^*} \Psi 
+ \big( \lambda e^{N X} + b \big)^{\frac{1}{N}} 
\cdot \del_i \del_{j^*} \Psi \; , \nonumber 
\end{align}
where $\del_{j^*} \Psi$ and $\del_i \del_{j^*} \Psi$ are given in
(\ref{deriv-Calabi}).

The metric of the submanifold of $\rho = 0$ ($d \rho = 0$) 
is obtained from (\ref{regular-Calabi}): 
\begin{align}
g_{i j^*} \big|_{\rho = 0} (\varphi, \varphi^*) 
\ &= \
b^{\frac{1}{N}} \del_i \del_{j^*} \Psi \; . \label{sub-mfd-Calabi}
\end{align}
Since $\Psi$ is the \kahler potential (\ref{CPN-Kahler}), 
$\del_i \del_{j^*} \Psi$ is nothing but 
the Fubini-Study metric (\ref{deriv-Calabi}) 
of ${\bf C}P^{N-1}$. 
Therefore we find that 
the total space is the complex line bundle 
over ${\bf C}P^{N-1}$ with fiber $\rho$.
This can be expected from the fact that 
there exists a Ricci-flat metric on the complex line bundle over 
any K\"{a}hler-Einstein manifolds~\cite{PP}. 

In the limit of $b\to 0$, this base manifold shrinks to 
zero-size and a singularity appears.
The \kahler potential (\ref{kahler-Calabi}) reduces to 
\begin{align}
{\cal K} \big|_{b = 0} 
\ &= \ 
\lambda^{\frac{1}{N}} |\sigma|^{2} \big( 1 + |\varphi^i|^2 \big) 
\ = \ \lambda^{\frac{1}{N}} \vec{\phi}^{\dagger} \vec{\phi}\;,
\end{align}
where $\phi^1 = \sigma$ and $\phi{}^i = \sigma \varphi^{i-1}$ 
($i = 2, 3, \cdots, N$). 
One might consider that 
this \kahler potential would give a flat metric, 
but it is not the case; 
We need a coordinate identification $\rho = \sigma^N /N$
(\ref{rho-Calabi}). 
The range of $\arg \rho$ has to be $0 \leq \arg \rho \leq 2 \pi$ to
avoid a conical singularity at $\rho = 0$ 
(in the case of $b \neq 0$),
that is,
the range of $\arg \sigma$ is $0 \leq \arg \sigma \leq 2 \pi/N$ and
all points with $\arg \sigma$ and $\arg \sigma + 2 \pi k/N$ ($k = 1,2,
\cdots, N$) are identified.
Therefore, the manifold of    
the singular limit is an orbifold ${\bf C}^N / {\bf Z}_N$. 
Imposing the parameter $b$ as a non-zero value,
we can replace the conical singularity with ${\bf C}P^{N-1}$ of
radius $b^{\frac{1}{2N}}$.
In the case of $N=2$, it is the Eguchi-Hanson 
space~\cite{EH}. 


\section{Conifold} \label{LQ}

In this section we discuss the gauge theoretical 
construction of the complex quadric surface
$Q^{N-2}= SO(N)/[SO(N-2) \times U(1)]$~\cite{HN1,HN2,HKNT}, 
followed by the discussion on 
the Ricci-flat metric on the conifold, 
which can be regarded as the complex line bundle 
over $Q^{N-2}$~\cite{HKN2}.

Imposing an appropriate constraint on the ${\bf C}P^{N-1}$ model, 
we can obtain the complex quadric surface
$Q^{N-2}$~\cite{HN1,HN2,HKNT}. 
The Lagrangian is given by 
\begin{align}
{\cal L} 
\ &= \ 
\int \! d^4 \theta
\big( \vec{\phi}^{\dagger} \vec{\phi} e^V - c V \big) 
+ \Big( 
\int \! d^2 \theta \, \phi_0 \vec{\phi}^T J \vec{\phi} + \mbox{c.c.} 
\Big) \; , \ \ \  \label{Lagrangian-Q}
\end{align}
where the \kahler potential 
is the same as the one of 
the ${\bf C}P^{N-1}$ model in (\ref{CPN}).
The integrand of the second term is 
the superpotential, in which 
$\phi_0 (x, \theta , \bar{\theta})$ is 
an auxiliary chiral superfield, 
and $J$ is the $SO(N)$-invariant 
rank-$2$ symmetric tensor, given by 
\begin{align}
J \ &= \ \left(
\begin{array}{ccc}
0 & {\bf 0} & 1 \\
{\bf 0} & {\bf 1}_{N-2} & {\bf 0} \\
1 & {\bf 0} & 0 
\end{array} \right) \; ,
\end{align}
where ${\bf 1}_{N-2}$ is an
$(N-2) \times (N-2)$ unit matrix.
By the integration over $V$, 
we obtain the \kahler potential (\ref{CPN-kahler-1}) 
in the same form with ${\bf C}P^{N-1}$. 
The integration over $\phi_0$ gives the constraint 
\begin{align}
\vec{\phi}^T J \vec{\phi} \ &= \ 0 \; . \label{F-con}
\end{align}
This can be solved as  
\begin{align}
\vec{\phi} \ &= \ \left(
\begin{array}{c}
1 \\
\varphi^i \\
- \half (\varphi^i)^2
\end{array} \right) \; , \label{QN-sol}
\end{align}
where we have chosen the gauge of $\phi^1 = 1$ 
using the complexified gauge symmetry, and 
$\varphi^i (x, \theta, \bar{\theta})$ are chiral superfields ($i
= 1, 2, \cdots, N-2$), 
whose scalar components parametrize the manifold. 
The \kahler potential written in terms of $\varphi^i$ is 
the one of $Q^{N-2}$ in the standard coordinates:  
\begin{align}
\Psi (\varphi, \varphi^{\dagger}) 
\ &= \ 
c \log \Big( 
1 + |\varphi^i|^2 + \frac{1}{4} (\varphi^i)^2 
(\varphi^{\dagger j})^2 
\Big) \; . \label{Q-Kahler}
\end{align}

Let us construct the Ricci-flat metric on the conifold 
as the complex line bundle 
over $Q^{N-2}$, as in the same manner as 
in the case of ${\bf C}P^{N-1}$.
We assume that the \kahler potential 
${\cal K}$ is a function of the
invariant superfield 
$X = \log \vec{\phi}^{\dagger} \vec{\phi}$:
\begin{align}
{\cal L} 
\ &= \ 
\int \! d^4 \theta \; 
{\cal K} (X) 
+ \Big( 
\int \! d^2 \theta \; \phi_0 \vec{\phi}^T J \vec{\phi} + \mbox{c.c.} 
\Big) \; . \label{LQ-L}
\end{align}
Since the holomorphic constraint from 
the integration over $\phi_0$ 
is again (\ref{F-con}), 
this non-compact manifold is a conifold. 
The constraint (\ref{F-con}) can be solved as
\begin{align}
\vec{\phi} \ &= \ \sigma \left(
\begin{array}{c}
1 \\
\varphi^i \\
- \half (\varphi^i)^2
\end{array} \right) \; ,
\end{align}
where $\sigma (x, \theta, \bar{\theta})$ 
is a chiral superfield.
By comparing this expression with (\ref{QN-sol}), 
we expect that $\sigma$ is 
a fiber, and $\varphi^i$ parametrize 
the base manifold $Q^{N-2}$, 
with the total space being the complex line bundle over $Q^{N-2}$. 
With this parametrization,
the invariant superfield $X$ is decomposed as
\bsubeq
\begin{align}
X \ &= \ 
\log |\sigma|^2 + \Psi \; , \ \ \ 
\Psi  \ = \ \log \xi \; , \\
\xi \ &\equiv \ 
1 + |\varphi^i|^2 + \frac{1}{4} (\varphi^{i})^2 
(\varphi^{\dagger j})^2 \; , 
\label{xi-q}
\end{align}
\esubeq
where $\Psi$ is the \kahler potential of 
$Q^{N-2}$ (\ref{Q-Kahler}).
This non-compact manifold can be 
regarded as ${\bf R} \times SO(N) / SO(N-2)$ at least locally.

As in the last section, we use the same letters for the chiral
superfields and their scalar components. 
The metric is defined in (\ref{metric})
where the holomorphic coordinates are given by 
$z^{\mu} = (\sigma , \varphi^i)$.
The determinant $\det g_{\mu \nu^*}$ is written as 
\begin{align}
\det g_{\mu \nu^*} 
\ &= \ 
\frac{1}{|\sigma|^2} {\cal K}'' 
\cdot \det_{i,j} 
\Big( 
{\cal K}' \frac{\del^2 X}{\del \varphi^i \del \varphi^{*j}} 
\Big) \; , 
\label{det-q0}
\end{align}
with the second derivative of $X$ being
\begin{align}
\frac{\del^2 X}{\del \varphi^i \del \varphi^{*j}} 
\ &= \ 
\del_i \del_{j^*} \Psi 
\ = \ 
\big( \delta_{i j} + \varphi^i \varphi^{*j} \big) \xi^{-1} 
- \big( \varphi^{*i} + \varphi^i (\varphi^{*k})^2 \big)
\big( \varphi^j + (\varphi^{l})^2 \varphi^{*j} \big) \xi^{-2} \;
. \label{deriv-q}
\end{align}
Under the complex isotropy transformation of $SO(N-2)^{\bf C}$, 
we can calculate the determinant in (\ref{det-q0}) with ease: 
\begin{align}
\det g_{\mu \nu^*} 
\ &= \ 
|\sigma|^{2N-6} e^{-(N-2)X} {\cal K}'' ({\cal K}')^{N-2} \; .  
\label{det-q}
\end{align}
The Ricci-flat condition (\ref{RF-condition}) 
can be solved for ${\cal K}'$:
\begin{align}
{\cal K}' \ &= \ \big( \lambda e^{(N-2) X} + b \big)^{\frac{1}{N-1}}
\; , \label{rf-sol-q}
\end{align}
where $\lambda$ is a constant and $b$ is an integration constant 
interpreted as a resolution parameter of the conical singularity.
Since it is sufficient for us to construct the metric from
(\ref{rf-sol-q}), 
we do not write down the explicit expression of the \kahler
potential itself (see \cite{HKN2}). 

We can immediately obtain the Ricci-flat metric from (\ref{rf-sol-q}).
The component $g_{\sigma \sigma^*}$ is given by
\begin{align}
g_{\sigma \sigma^*} 
\ &= \ 
\frac{N-2}{N-1} \lambda 
\big( \lambda e^{(N-2) X} + b \big)^{-\frac{N-2}{N-1}} 
e^{(N-2) \Psi} |\sigma|^{2N-6} \; .
\end{align}
This component is not regular at the $\sigma = 0$ surface: $g_{\sigma
\sigma^*} |_{\sigma = 0} = 0$.
However this is just a coordinate singularity of $z^{\mu} =
(\sigma , \varphi^i)$.
As in the case of the line bundle over ${\bf C}P^{N-1}$, 
we perform the following transformation:
\begin{align}\rho \ &\equiv \ \frac{\sigma^{N-2}}{N-2} \; ,
\end{align}
with $\varphi^i$ being unchanged.
Under this transformation, 
the metric in this coordinate system becomes regular in the whole region
(including $\rho = 0$):
\bsubeq
\begin{align}
g_{\rho \rho^*} \ &= \ \lambda \frac{N-2}{N-1} 
 \big( \lambda e^{(N-2) X} + b \big)^{\frac{2-N}{N-1}} e^{(N-2) \Psi}
 \; , \\
g_{\rho j^*} \ &= \ \lambda \frac{(N-2)^2}{N-1} 
 \big( \lambda e^{(N-2) X} + b \big)^{\frac{2-N}{N-1}} e^{(N-2) \Psi} 
 \rho^* \cdot \del_{j^*} \Psi \; , \\
g_{i j^*} \ &= \ \lambda \frac{(N-2)^3}{N-1} 
 \big( \lambda e^{(N-2) X} + b \big)^{\frac{2-N}{N-1}} e^{(N-2) \Psi} 
 |\rho|^2 \cdot \del_i \Psi \del_{j^*} \Psi \nonumber \\
\ & \ \ \ \ 
 + \big( \lambda e^{(N-2) X} + b \big)^{\frac{1}{N-1}} 
 \cdot \del_i \del_{j^*} \Psi \; . 
\end{align}
\esubeq
The metric of the submanifold of $\rho = 0$ ($d \rho = 0$) 
is obtained as
\begin{align}
\dps g_{i j^*} \big|_{\rho = 0} (\varphi, \varphi^*) 
\ &= \
b^{\frac{1}{N-1}} \del_i \del_{j^*} \Psi \; .
\end{align}
This is the metric of $Q^{N-2}$ given in (\ref{deriv-q}), 
since $\Psi$ is its \kahler potential (\ref{Q-Kahler}).
In the limit of $b \to 0$, 
this submanifold shrinks to zero-size and the total space becomes a
conifold.
When $b\neq 0$, 
the conical singularity is resolved by $Q^{N-2}$ of radius
$b^{\frac{1}{2(N-1)}}$. 
Thus this non-compact Calabi-Yau manifold can be 
regarded as the complex line
bundle over $Q^{N-2}$ with fiber $\rho$.

Let us make some comments.
When $N=3$, the line bundle over $Q^1$ coincides with the 
Eguchi-Hanson gravitational instanton.
In the case of $N=4$,
the manifold becomes the line bundle over $Q^2 \simeq S^2 \times S^2$
(the radii of these two $S^2$ coincide)~\cite{PT}.
The way of removing this conical singularity 
is different from either the
deformation by $S^3$~\cite{conifold,conifold2} 
or the small resolution by $S^2$~\cite{conifold} known 
in the six-dimensional conifold.


\section{Conifolds with Isometries of $E_6$ and $E_7$} \label{LE}

In the last section 
we have explained the $O(N)$ symmetric conifold which can be 
regarded as the complex line bundle over $Q^{N-2}$. 
In this section, 
we construct the {\it new} conifolds whose isometries are
the exceptional groups $E_6$ and $E_7$, 
in the same way as the previous sections. 


\subsection{Hermitian Symmetric Spaces $E_6/[SO(10)\times U(1)]$ 
and $E_7/[E_6\times U(1)]$}

In this subsection
we give the gauge theoretical construction of 
the HSS of the exceptional groups, 
$E_6/[SO(10) \times U(1)]$ and
$E_7 / [E_6 \times U(1)]$~\cite{HN1}.
We start from the Lagrangian
\begin{align}
{\cal L} 
\ &= \ 
\int \! d^4 \theta \; 
(\vec{\phi}^{\dagger} \vec{\phi} e^V - c \, V ) 
+ \Big( \int \! d^2 \theta \; W (\vec{\phi}_0, \vec{\phi}) 
+ \mbox{c.c.} \Big) \; , \label{Lagrangian-E}
\end{align}
where $\vec{\phi} (x, \theta, \bar{\theta})$ are 
chiral superfields belonging to the fundamental
representation of the global symmetry $E_6$ or $E_7$, 
and  $V (x, \theta, \bar{\theta})$ is an auxiliary
vector superfield of $U(1)_{\rm local}$ gauge symmetry.
Later we define the superpotential $W$ as a function of    
$\vec{\phi}$ and auxiliary chiral superfields  
$\vec{\phi}_0 (x, \theta, \bar{\theta})$, which also belong to 
the fundamental representation. 

\ul{\sl $E_6 / [SO(10) \times U(1)]$}. \hfil\break
We consider the global symmetry 
$G = E_6$ and a $U(1)_{\rm local}$ gauge symmetry.
The chiral superfields $\vec{\phi}$ belong to the fundamental 
representation ${\bf 27}$ of $E_6$.
We decompose $E_6$ under 
its maximal subgroup $SO(10) \times U(1)$.  

Since the fundamental representation can be decomposed as 
${\bf 27} = ({\bf 1},4) \oplus ({\bf 16},1) 
\oplus ({\bf 10},-2)$~\cite{Sl}, where the second entries 
are the $U(1)$ charges, 
the fields $\vec{\phi}$ can be written as
\begin{align}
\vec{\phi} \ &= \ \begin{pmatrix}
x \cr
y_{\alpha} \cr
z^A
\end{pmatrix} \; .\label{decom.E6}
\end{align}
Here, $x$, $y_{\alpha}$ ($\alpha=1,\cdots,16$) and 
$z^A$ ($A=1,\cdots,10$) are an $SO(10)$ scalar, 
a Weyl spinor and a vector, respectively.  

The decomposition of the tensor product, 
${\bf 27} \otimes {\bf 27} = {\bf \ol{27}}_{\rm s} \oplus \cdots$, 
implies that there exist the rank-$3$ symmetric 
invariant tensor $\Gamma_{ijk}$ and 
its complex conjugate $\Gamma^{ijk}$, 
whose components can be read from the invariant~\cite{KS}
\begin{align}
 I_3 \ &\equiv \ \Gamma_{ijk} \phi^i \phi^j \phi^k  
   \ = \ x z^2 + \frac{1}{\sqrt{2}} 
   z^A ( y C \sigma_A^{\dagger} y ) \; ,\label{I3}
\end{align}
where we have used the decomposition (\ref{decom.E6}).
Here the indices $i,j,k$ run from 1 to 27.
For the product of the invariant tensors, 
some identities 
\bsubeq
\begin{align}
 \Gamma_{ijk}\Gamma^{ijl} \ &= \ 10 \delta_k^l 
 \; , \label{E6-id.} \\
 \Gamma_{ijk}(\Gamma^{jl\{m}\Gamma^{np\}k}) 
\ &= \ {\delta_i}^{\{l} \Gamma^{mnp\}} \; , \label{Springer}
\end{align}
\esubeq
hold, where we have used the notation 
$A^{\{ij\cdots\}}=A^{ij\cdots} + A^{ji\cdots} + \cdots$.
(The second one is called the Springer relation.)
These identities are used many times 
in the analysis of the $E_7$ algebra.

Then we define the superpotential by  
\begin{align}
 W (\vec{\phi}_0, \vec{\phi}) 
\ &= \ 
\Gamma_{ijk} {\phi_0}^i \phi^j \phi^k \; ,
\end{align}
where $\vec{\phi}_0$ represent auxiliary fields whose 
$U(1)_{\rm local}$ charge is $-2$ when  
the $U(1)_{\rm local}$ charge of $\vec{\phi}$ is $1$. 
The equations of motion for the auxiliary fields ${\phi_0}^i$, 
$\del W /\del \phi_0{}^i = \Gamma_{ijk} \phi^j \phi^k = 0$, 
are 
\bsubeq
\label{E6_F-flat}
\begin{align}
\del W / \del {z_0}^A 
\ &= \ 2 z_A x + \frac{1}{\sqrt{2}} ( y C \sigma_A^{\dagger} y ) \ = \
0 \; , \\ 
\del W / \del y_{0\alpha} 
\ &= \ \sqrt{2} (C \sigma_A^{\dagger} y)^{\alpha} z^A  
\ = \ 0 \;, \\
\del W / \del x_0 
\ &= \ z^2 \ = \ 0 \; ,
\end{align}
\esubeq
In the second equation, we have used the fact that 
$(C \sigma_A^{\dagger})^{\alpha\beta}$ is symmetric.
The first equation can be solved to yield
\begin{align}
z_A \ &= \ - \frac{1}{2 \sqrt{2} x} ( y C \sigma_A^{\dagger} y) \; . 
\label{E6-constraint}
\end{align}
We can show that the last two equations 
are not independent of the first~\cite{HN1}. 
Therefore we can write the fundamental field as
\begin{align}
\vec{\phi} \ &= \ \left(
\begin{array}{c}
x \\
y_{\alpha} \\
- \frac{1}{2 \sqrt{2} x}(y C \sigma_A^{\dagger} y)
\end{array} \right) \; . \label{csf-e6-1}
\end{align}
Integrating out the auxiliary field $V$,
we obtain the \kahler potential as the same form of
(\ref{CPN-kahler-1}).
Using complexified gauge symmetry, 
we can choose the gauge of $x=1$: 
\begin{align}
\vec{\phi} \ &= \ \left(
\begin{array}{c}
1 \\
\varphi_{\alpha} \\
- \frac{1}{2 \sqrt{2}}(\varphi C \sigma_A^{\dagger} \varphi)
\end{array} \right) \; ,
\end{align}
where we have rewritten $y_{\alpha}$ as $\varphi_{\alpha}$.  
Substituting this into (\ref{CPN-kahler-1}), we obtain the
following expression:
\begin{align}
\Psi (\varphi, \varphi^{\dagger}) 
\ &= \ 
c \log 
\Big( 1 + |\varphi_{\alpha}|^2 
+ \frac{1}{8} | \varphi C \sigma_A^{\dagger} \varphi |^2 \Big) \; . 
\label{E6-Kahler}
\end{align}
This is the \kahler potential of $E_6/[SO(10)\times U(1)]$ 
in the standard coordinates.  


\ul{\sl $E_7 / [ E_6 \times U(1)]$}. \hfil\break
The global symmetry in this case is 
$G = E_7$ and the local symmetry is $U(1)_{\rm local}$.
The chiral superfields $\vec{\phi}$ belong to the fundamental 
representation ${\bf 56}$ of $E_7$. 
Since the maximal subgroup of $E_7$ is $E_6 \times U(1)$, 
we construct $E_7$ from $E_6 \times U(1)$. 

The fundamental representation can be decomposed as
${\bf 56} = ({\bf 27}, -\frac{1}{3})\oplus ({\bf \ol{27}},\frac{1}{3}) 
\oplus ({\bf 1},-1) \oplus ({\bf 1},1)$. 
Thus we write $\vec{\phi}$ as 
\begin{align}
 \vec{\phi} \ &= \ \left(
\begin{array}{c}
x \\
y^i \\ 
z_i \\
 w
\end{array} \right) \; ,  \label{decom.E7}
\end{align}
where $y^i$ and $z_i$ are ${\bf 27}$ and ${\bf \ol{27}}$
representations of $E_6$, 
respectively; $x$ and $w$ are scalars.

There exists 
the rank-$4$ symmetric invariant tensor 
$d_{\alpha \beta \gamma \delta}$, 
whose components can be read from the invariant~\cite{HN1}
\begin{align}
 I_4 \ &\equiv \ d_{\alpha \beta \gamma \delta} 
     \phi^{\alpha} \phi^{\beta} \phi^{\gamma} \phi^{\delta} \nonumber \\
\ &= \ -\half (x w - y^i z_i)^2 
-\frac{1}{3} w \Gamma_{ijk} y^i y^j y^k 
- \frac{1}{3} x \Gamma^{ijk} z_i z_j z_k 
+ \half \Gamma^{ijk} \Gamma_{ilm} z_j z_k y^l y^m \; ,\label{I4}
\end{align}
where $I_4$ is invariant due to the Springer relation 
for the $E_6$ invariant tensor (\ref{Springer}). 

By using this invariant tensor,
the superpotential invariant under 
$E_7 \times U(1)_{\rm local}$ is given by 
\begin{align}
W (\vec{\phi}_0, \vec{\phi}) 
\ &= \ d_{\alpha \beta \gamma \delta} 
  {\phi_0}^{\alpha} \phi^{\beta} \phi^{\gamma} \phi^{\delta} \; ,
\end{align}
where $\phi_0{}^{\alpha}$ are auxiliary fields 
belonging to $({\bf 56},-3)$. 
Here the second component is the $U(1)_{\rm local}$ charge 
assigned to cancel the  $U(1)_{\rm local}$ charge of $\phi^{\alpha}$.
The integration over $\phi_0{}^{\alpha}$ gives 
the constraints $\del W / \del \phi_0{}^{\alpha} = 
d_{\alpha \beta \gamma \delta} 
\phi^{\beta} \phi^{\gamma} \phi^{\delta} = 0$:
\bsubeq \label{E7_F-flat}
\begin{align}
\del W / \del {y_0}^i
\ &= \ w (x z_i - \Gamma_{ijk} y^j y^k) - z_i y^j z_j
  + \Gamma^{jkl} \Gamma_{jim} z_k z_l y^m \ = \ 0 \; , \\
\del W / \del w_0
\ &= \ x y^i z_i - w x^2 -\frac{1}{3} \Gamma_{ijk} y^i y^j y^k =
0 \; , \\
\del W / \del z_{0 i} 
\ &= \ x (w y^i - \Gamma^{ijk} z_j z_k) - y^i y^j z_j
  + \Gamma^{jik} \Gamma_{jlm} z_k y^l y^m \ = \ 0 \; , \\
\del W / \del x_0 
\ &= \ w y^i z_i - x w^2 -\frac{1}{3} \Gamma^{ijk} z_i
z_j z_k \ = \ 0 \; . 
\end{align}
\esubeq
First two equations can be solved as~\cite{HN1}
\begin{align}
z_i \ &= \ 
\frac{1}{2 x} \Gamma_{ijk} y^j y^k \; , \ \ \ 
 w \ = \ 
\frac{1}{6 x^2} \Gamma_{ijk} y^i y^j y^k \; . 
\label{E7-ans.}
\end{align}
It was shown in \cite{HN1} 
that the last two equations are not independent.
Then, we have 
\begin{align}
 \vec{\phi} \ &= \ \left(
\begin{array}{c}
x \\
y^i \\
\frac{1}{2x} \Gamma_{ijk} y^j y^k \\
\frac{1}{6x^2} \Gamma_{ijk} y^i y^j y^k
\end{array} \right) \; . \label{E7_F-flat2}
\end{align}
We obtain the \kahler potential (\ref{CPN-kahler-1}) from 
the integration over $V$.
The complexified gauge symmetry can be fixed as $x=1$: 
\begin{align}
\vec{\phi} \ &= \ \left(
\begin{array}{c}
1 \\
\varphi^i \\
\frac{1}{2} \Gamma_{ijk} \varphi^j \varphi^k \\
\frac{1}{6} \Gamma_{ijk} \varphi^i \varphi^j \varphi^k
\end{array} \right) \; ,
\end{align}
where we have rewritten $y^i$ as $\varphi^i$. 
Substituting this into (\ref{CPN-kahler-1}),
we obtain 
\begin{align}
\Psi (\varphi, \varphi^{\dagger}) 
\ &= \ 
c \log 
\Big( 1 + |\varphi^i|^2 + \frac{1}{4} |\Gamma_{i j k} \varphi^j
\varphi^k|^2 
+ \frac{1}{36} |\Gamma_{i j k} \varphi^i \varphi^j \varphi^k |^2 \Big)
\; .  \label{E7-Kahler}
\end{align}
This is the \kahler potential of $E_7/[E_6\times U(1)]$ 
in the standard coordinates. 


\subsection{Construction of Line Bundles}

In this subsection,
we construct the conifolds with isometries of $E_6$ and $E_7$, 
which can be regarded as the complex line
bundles over $E_6 / [SO(10) \times U(1)]$ 
and $E_7 / [E_6 \times U(1)]$, respectively.
As in the previous sections,
we obtain these manifolds when $U(1)_{\rm local}$ is not gauged.

First we discuss the case of $E_6 / [SO(10) \times U(1)]$.
Since the $U(1)_{\rm local}$ symmetry is not gauged,
the chiral superfield satisfying only the F-term constraints 
(\ref{E6_F-flat}) can be written as (\ref{non-cpt-csf}):
\begin{align}
\vec{\phi} \ &= \ \sigma \left(
\begin{array}{c}
1 \\
\varphi_{\alpha} \\
- \frac{1}{2 \sqrt{2}} (\varphi C \sigma_A^{\dagger} \varphi)
\end{array} \right) \; , \label{non-cpt-e6}
\end{align}
where $\varphi_{\alpha} (x , \theta , \bar{\theta})$ are chiral
superfields belonging to an $SO(10)$ Weyl spinor 
($\alpha = 1 ,2, \cdots, 16$), 
and $\sigma (x, \theta, \bar{\theta})$ 
is a chiral superfield.
We find that $\sigma$ and $\varphi_{\alpha}$ parametrize a fiber and 
a base manifold, 
with the total space being a complex line bundle over $E_6 / [SO(10)
\times U(1)]$.
$\sigma_A$ are $SO(10)$ $\gamma$-matrices in the Weyl spinor
basis ($A = 1,2, \cdots, 10$) and $C$ is a charge conjugation matrix
(represented in appendix \ref{SO10}).

Under the expression (\ref{non-cpt-e6}), 
the invariant superfield $X = \log \vec{\phi}^{\dagger} \vec{\phi}$
is decomposed as
\bsubeq \label{xi-e6}
\begin{align}
X 
\ &= \
\log |\sigma|^2 + \Psi \; , \ \ \ 
\Psi \ = \ \log \xi \; , \\
\xi \ &\equiv \ 
1 + |\varphi_{\alpha}|^2 
+ \frac{1}{8} |\varphi C \sigma_A^{\dagger} \varphi|^2 \; ,
\end{align}
\esubeq
where $\Psi$ is the \kahler potential 
of $E_6 / [SO(10) \times U(1)]$ defined in (\ref{E6-Kahler}).
This non-compact manifold can be regarded as 
${\bf R} \times E_6 / SO(10)$ at least locally.

In the case of $E_7 / [E_6 \times U(1)]$, 
the chiral superfields satisfying only the F-term constraints 
(\ref{E7_F-flat}) can be written in the same way 
as (\ref{non-cpt-e6}):
\begin{align}
\vec{\phi} \ &= \ \sigma \left(
\begin{array}{c}
1 \\
\varphi^i \\
\frac{1}{2} \Gamma_{ijk} \varphi^j \varphi^k \\
\frac{1}{6} \Gamma_{ijk} \varphi^i \varphi^j \varphi^k
\end{array} \right) \; . \label{non-cpt-e7}
\end{align}
Here chiral superfields 
$\varphi^i (x , \theta , \bar{\theta})$, 
belonging to the ${\bf 27}$ representation of $E_6$,
parametrize a base manifold, 
and $\sigma (x , \theta, \bar{\theta})$ is a chiral superfield 
parametrizing a fiber, 
with the total space being 
a complex line bundle over $E_7 / [E_6 \times U(1)]$. 
$\Gamma_{ijk}$ is the rank-3 symmetric tensor invariant of $E_6$. 
Under the expression (\ref{non-cpt-e7}), 
the invariant superfield $X= \log \vec{\phi}^{\dagger} \vec{\phi}$
is decomposed as
\bsubeq \label{xi-e7}
\begin{align}
X \ &= \ 
\log |\sigma|^2 + \Psi \; , \ \ \ 
\Psi \ = \ \log \xi \; , \\
\xi \ &\equiv \ 
1 + |\varphi^i|^2 
+ \frac{1}{4} |\Gamma_{ijk} \varphi^j \varphi^k|^2 
+ \frac{1}{36} |\Gamma_{ijk} \varphi^i \varphi^j \varphi^k|^2 \; ,
\end{align}
\esubeq
where $\Psi$ is the \kahler potential of 
$E_7 / [E_6 \times U(1)]$ given by (\ref{E7-Kahler}).
This manifold can be also regarded as ${\bf R} \times E_7 / E_6$, 
at least locally.

The metric is defined in (\ref{metric})
where the holomorphic coordinates (and their conjugates) are 
$z^{\mu} = (\sigma , \varphi_{\alpha})$ 
[$z^{* \mu} = (\sigma^* , \varphi^*_{\alpha})$] 
for $E_6 / [SO(10) \times U(1)]$ or 
$z^{\mu} = (\sigma , \varphi^i)$ [$z^{*\mu} = (\sigma^* , \varphi^*_i)$] 
for $E_7 / [E_6 \times U(1)]$.
(The notation of $\varphi^*_i$ is because of the fact that 
$\varphi^i$ belong to a complex representation of $E_6$.)
The determinant $\det g_{\mu \nu^*}$ is written as 
\begin{align}
\det g_{\mu \nu^*} \ &= \ \left\{
\begin{array}{l@{\ls}l}
\dps \frac{1}{|\sigma|^2} {\cal K}'' \cdot \det_{\alpha , \beta} \Big(
{\cal K}' \frac{\del^2 X}{\del \varphi_{\alpha} \del \varphi^*_{\beta}}
\Big) & \mbox{for $E_6 / [SO(10) \times U(1)]$} \; , \\
\dps \frac{1}{|\sigma|^2} {\cal K}'' \cdot \det_{i , j} 
\Big( {\cal K}' \frac{\del^2 X}{\del \varphi^i \del \varphi^*_j}
\Big) & \mbox{for $E_7 / [E_6 \times U(1)]$} \; . \\
\end{array} \right. \label{det-e0}
\end{align}
The $X$ differentiated once or twice are calculated as
\bsubeq
\begin{align}
\Pd{X}{\varphi_{\alpha}} 
\ &= \ 
\del_{\alpha} \Psi 
\ = \ 
\frac{1}{\xi} \Big\{ 
\varphi^*_{\alpha}
+ \frac{1}{4} (C \sigma_A^{\dagger} \varphi)^{\alpha} 
(\varphi^* \sigma^A C^{\dagger} \varphi^*)
\Big\} \; , \\ 
\frac{\del^2 X}{\del \varphi_{\alpha} \del \varphi^*_{\beta}} 
\ &= \ 
\del_{\alpha} \del_{\beta^*} \Psi 
\ = \ 
\frac{1}{\xi} \Big\{ 
\delta_{\alpha \beta} 
+ \half ( \sigma^A C^{\dagger} \varphi^*)^{\beta} 
(C \sigma_A^{\dagger} \varphi)^{\alpha} 
\Big\} \nonumber \\
& \LS \LS  - \frac{1}{\xi^2} \Big\{ \varphi^*_{\alpha}
+ \frac{1}{4} ( C \sigma_A^{\dagger} \varphi )^{\alpha} 
( \varphi^* \sigma^A C^{\dagger} \varphi^*) \Big\} 
\Big\{ \varphi_{\beta} + \frac{1}{4} 
( \sigma^A C^{\dagger} \varphi^* )^{\beta} 
( \varphi C \sigma_A^{\dagger} \varphi) \Big\} \; , 
\label{deriv-e6}
\end{align}
\esubeq
for $E_6 / [SO(10) \times U(1)]$ or
\bsubeq
\begin{align}
\Pd{X}{\varphi^i} 
\ &= \ 
\del_i \Psi 
\ = \ 
\frac{1}{\xi} 
\Big\{ \varphi^*_i 
+ \half (\Gamma_{i j k} \varphi^k) (\Gamma^{j l m} \varphi^*_l
\varphi^*_m) + \frac{1}{12} (\Gamma_{i j k} \varphi^j \varphi^k)
(\Gamma^{l m n} \varphi^*_l \varphi^*_m \varphi^*_n) \Big\} \; , \\
\frac{\del^2 X}{\del \varphi^i \del \varphi^*_j} 
\ &= \ 
\del_i \del_{j^*} \Psi 
\ = \ 
\frac{1}{\xi} \Big\{ \delta_{i j} 
+ (\Gamma_{i k l} \varphi^l) (\Gamma^{j k m} \varphi^*_m)
+ \frac{1}{4} (\Gamma_{i k l} \varphi^{k} \varphi^{l})
(\Gamma^{j m n} \varphi^*_m \varphi^*_n) \Big\} \nonumber \\
& \LS \ls \ \ \ - \frac{1}{\xi^2} \Big\{ \varphi^*_i
+ \half (\Gamma_{i k l} \varphi^l) (\Gamma^{k m n} \varphi^*_m
\varphi^*_n) 
+ \frac{1}{12} (\Gamma_{i k l} \varphi^k \varphi^l)
(\Gamma^{m n p} \varphi^*_m \varphi^*_n \varphi^*_p) \Big\}
\nonumber \\
& \LS \LS \times \Big\{ \varphi^j 
+ \half (\Gamma_{k l m} \varphi^l \varphi^m) (\Gamma^{j k n} \varphi^*_n)
+ \frac{1}{12} (\Gamma_{k l m} \varphi^k \varphi^l \varphi^m)
(\Gamma^{j n p} \varphi^*_n \varphi^*_p) \Big\} \; ,
\label{deriv-e7}
\end{align}
\esubeq
for $E_7 / [E_6 \times U(1)]$.

Using the complex isotropy transformation of $SO(10)^{\bf C}$ 
[$E_6{}^{\bf C}$],
we can put one component of 
$\varphi_{\alpha}$ [$\varphi^i$] to non-zero 
with others being zero values, 
as a vacuum expectation value.
This is because 
the determinant $\det g_{\mu \nu^*}$ is invariant
under this transformation.
Thus the partial differential equation (\ref{RF-condition})
reduces to an ordinary differential equation.
Here we put $\varphi_1 \neq 0$ [$\varphi^1 \neq 0$], and others to zero:
\bsubeq
\begin{align}
\varphi_1 \ &\neq \ 0 \; , \ \ \ 
\varphi_2 \ = \ \varphi_3 \ = \ \cdots \ = \ \varphi_{16} \ = \ 0 
&& \mbox{for $E_6 / [SO(10) \times U(1)]$} \; , \\
\varphi^1 \ &\neq \ 0 \; , \ \ \ 
\varphi^2 \ = \ \varphi^3 \ = \ \cdots \ = \ \varphi^{27} \ = \ 0 
&& \mbox{for $E_7 / [E_6 \times U(1)]$} \; .
\end{align}
\esubeq
Under the representations (\ref{pauli}) and (\ref{cc-matrix}), 
we calculate second derivatives (\ref{deriv-e6}) and (\ref{deriv-e7}):
\bsubeq
\begin{align}
\frac{\del^2 X}{\del \varphi_{\alpha} \del \varphi^*_{\beta}} \ &= \
\left\{
\begin{array}{ll}
\xi^{-2} & \ \ \ \alpha = \beta = 1 \\
\xi^{-1} & \ \ \ \alpha = \beta = 2, 3, 4, 5, 10, 11, 12, 14, 15, 16 \\
1 & \ \ \ \alpha = \beta = 6, 7, 8, 9, 13 \\
0 & \ \ \ \mbox{otherwise} 
\end{array} \right. && \mbox{for $E_6 / [SO(10) \times U(1)]$} \; , \\
\frac{\del^2 X}{\del \varphi^i \del \varphi^*_j} \ &= \ \left\{
\begin{array}{ll}
\xi^{-2} & \ \ \ i = j = 1 \\
\xi^{-1} & \ \ \ i = j = 2, 3, \cdots, 17 \\ 
1 & \ \ \ i = j = 18,19, \cdots , 27 \\
0 & \ \ \ \mbox{otherwise}
\end{array} \right. && \mbox{for $E_7 / [E_6 \times U(1)]$} \; .
\end{align}
\esubeq
Then the determinant is obtained as
\begin{align}
\det g_{\mu \nu^*} \ &= \ \left\{
\begin{array}{l@{\ls}l}
\dps |\sigma|^{22} e^{-12X} {\cal K}'' ({\cal K}')^{16} & \mbox{for
$E_6 / [SO(10) \times U(1)]$} \; , \\ 
\dps |\sigma|^{34} e^{-18X} {\cal K}'' ({\cal K}')^{27} &
\mbox{for $E_7 / [E_6 \times U(1)]$} \; ,
\end{array} \right. \label{det-e6-e7}
\end{align}
where we have used $\xi = |\sigma|^{-2} e^X$ from 
(\ref{xi-e6}) or (\ref{xi-e7}).
The Ricci-flat condition (\ref{RF-condition}) 
can be solved for ${\cal K}'$:
\begin{align}
{\cal K}' \ &= \ \left\{
\begin{array}{l@{\ls}l}
\dps \big( \lambda e^{12 X} + b \big)^{\frac{1}{17}} 
& \mbox{for $E_6 / [SO(10) \times U(1)]$} \; , \\ 
\dps \big( \lambda e^{18 X} + b \big)^{\frac{1}{28}} 
& \mbox{for $E_7 / [E_6 \times U(1)]$} \; ,
\end{array} \right. \label{rf-sol-e}
\end{align}
where $\lambda$ is a constant and $b$ is an integration constant
regarded as a resolution parameter of the conical singularity.
The \kahler potential can be also 
calculated using (\ref{fn-I}), to give
\begin{align}
{\cal K} (X) \ &= \ \left\{
\begin{array}{l@{\ls}l}
\dps \frac{17}{12} \Big[ 
\big( \lambda e^{12 X} + b \big)^{\frac{1}{17}} 
+ b^{\frac{1}{17}} \cdot I \big( b^{- \frac{1}{17}}
\big( \lambda e^{12 X} + b \big)^{\frac{1}{17}} ; 17 \big) \Big] & 
\mbox{for $E_6 / [SO(10) \times U(1)]$} \; , \\ 
\\
\dps \frac{14}{9} \Big[ 
\big( \lambda e^{18 X} + b \big)^{\frac{1}{28}} 
+ b^{\frac{1}{28}} \cdot I \big( b^{- \frac{1}{28}}
\big( \lambda e^{18 X} + b \big)^{\frac{1}{28}} ; 28 \big) \Big] &
\mbox{for $E_7 / [E_6 \times U(1)]$} \; .
\end{array} \right. \label{kahler-sol-e}
\end{align}

Now we can immediately obtain the Ricci-flat metric from 
the solution (\ref{rf-sol-e}) or (\ref{kahler-sol-e}).
The component $g_{\sigma \sigma^*}$ is calculated as
\begin{align}
g_{\sigma \sigma^*} \ &= \ \left\{
\begin{array}{ll}
\dps \frac{12}{17} \lambda 
\big( \lambda e^{12 X} + b \big)^{-\frac{16}{17}} 
e^{12 \Psi} |\sigma|^{22} 
& \dps \ls \mbox{for} \ \ E_6 / [SO(10) \times U(1)] \; , \\
\\
\dps \frac{9}{14} \lambda 
\big( \lambda e^{18 X} + b \big)^{-\frac{27}{28}} 
e^{18 \Psi} |\sigma|^{34} 
& \dps \ls \mbox{for} \ \ E_7 / [E_6 \times U(1)] \; . 
\end{array} \right.
\end{align}
Each is not regular at the $\sigma = 0$ surface: 
$g_{\sigma \sigma^*} |_{\sigma = 0} = 0$.
However it is just a coordinate singularity; 
If we perform the coordinate transformation
\begin{align}
\rho \ &\equiv \ \left\{
\begin{array}{l@{\ls}l}
\sigma^{12}/{12} & \mbox{for $E_6 / [SO(10) \times U(1)]$} \; , \\
\sigma^{18}/{18} & \mbox{for $E_7 / [E_6 \times U(1)]$} \; ,
\end{array} \right. 
\end{align}
with $\varphi_{\alpha}$ or $\varphi^i$ being unchanged, 
the components of the metrics in new coordinates become
\begin{align}
g_{\rho \rho^*} 
\ &= \ 
\frac{12}{17} \lambda 
\big( \lambda e^{12 X} + b \big)^{- \frac{16}{17}} e^{12 \Psi} \; ,
\nonumber \\
g_{\rho \beta^*} 
\ &= \ 
\frac{144}{17} \lambda 
\big( \lambda e^{12 X} + b \big)^{- \frac{16}{17}} 
e^{12 \Psi} \rho^* \cdot \del_{\beta^*} \Psi \; , \\
g_{\alpha \beta^*} 
\ &= \ 
\frac{1728}{17} \lambda 
\big( \lambda e^{12 X} + b \big)^{- \frac{16}{17}} 
e^{12 \Psi} |\rho|^2 \cdot \del_{\alpha} \Psi \del_{\beta^*} \Psi 
+ \big( \lambda e^{12 X} + b \big)^{\frac{1}{17}} 
\cdot \del_{\alpha} \del_{\beta^*} \Psi \; , 
\nonumber
\end{align}
for $E_6 / [ SO(10) \times U(1)]$, and
\begin{align}
g_{\rho \rho^*} 
\ &= \ 
\frac{9}{14} \lambda 
\big( \lambda e^{18 X} + b \big)^{- \frac{27}{28}} e^{18 \Psi} \; , 
\nonumber \\
g_{\rho j^*} 
\ &= \ 
\frac{81}{7} \lambda 
\big( \lambda e^{18 X} + b \big)^{- \frac{27}{28}} 
e^{18 \Psi} \rho^* \cdot \del_{j^*} \Psi \; , \\
g_{i j^*} 
\ &= \ 
\frac{1458}{7} \lambda 
\big( \lambda e^{18 X} + b \big)^{- \frac{27}{28}} 
e^{18 \Psi} |\rho|^2 \cdot \del_{i} \Psi \del_{j^*} \Psi 
+ \big( \lambda e^{18 X} + b \big)^{\frac{1}{28}} 
\cdot \del_{i} \del_{j^*} \Psi \; ,
\nonumber 
\end{align}
for $E_7 / [E_6 \times U(1)]$.
Hence, each metric is regular in the whole region (including $\rho = 0$).

The metric of each submanifold of 
$\rho = 0$ ($d \rho = 0$) is 
\bsubeq
\begin{align}
g_{\alpha \beta^*} \big|_{\rho = 0} (\varphi, \varphi^*) 
\ &= \
b^{\frac{1}{17}} \del_{\alpha} \del_{\beta^*} \Psi  
&& \mbox{for $E_6 / [SO(10) \times U(1)]$} \; , \\
g_{i j^*} \big|_{\rho = 0} (\varphi, \varphi^*) 
\ &= \
b^{\frac{1}{28}} \del_i \del_{j^*} \varphi 
&& \mbox{for $E_7 / [ E_6 \times U(1)]$} \; .
\end{align}
\esubeq
Since each $\Psi$ is the \kahler potential 
of $E_6/[SO(10) \times U(1)]$ or $E_7 / [E_6 \times U(1)]$, 
$\del_{\alpha} \del_{\beta^*} \Psi$ or 
$\del_i \del_{j^*} \Psi$ is the metric of this manifold, 
given in (\ref{deriv-e6}) and (\ref{deriv-e7}).
Therefore we find that the total space is 
the complex line bundle 
over $E_6/[SO(10) \times U(1)]$ 
or $E_7 / [E_6 \times U(1)]$ with a fiber $\rho$.
In the limit of $b \to 0$,
each submanifold shrinks to zero-size and 
the conical singularity appears. 
Each conical singularity is resolved by $E_6 / [SO(10) \times U(1)]$
or $E_7 / [E_6 \times U(1)]$ of radius $b^{\frac{1}{34}}$ or
$b^{\frac{1}{56}}$.


\section{Non-compact Calabi-Yau Manifolds from 
Non-abelian Gauge Theories} \label{LG}

In this section we review the construction of 
non-compact Calabi-Yau manifolds 
using the non-abelian gauge theories, 
which provides other generalizations of 
the complex line bundle over ${\bf C}P^{N-1}$~\cite{HKN3}.
First we construct the Grassmann manifold
$G_{N,M}$ using non-abelian gauge theory~\cite{Ao,HKLR}.  
Imposing holomorphic constraints on $G_{2N,N}$, 
we obtain the rests of HSS, 
$SO(2N)/U(N)$ and $Sp(N)/U(N)$~\cite{HN1}.
Second we study the non-compact Calabi-Yau manifolds 
as the complex line bundles 
over these compact manifolds in detail.

Let $\Phi (x, \theta, \bar{\theta})$ be an $N \times M$ matrix-valued
chiral superfield and 
$V (x, \theta, \bar{\theta}) = V^A T_A$ be 
a vector superfield taking a value in 
the Lie algebra of $U(M)$.
The global symmetry $SU(N)$ acts on $\Phi$ from the left:  
$\Phi \ \to \ \Phi' \ = \ g \Phi$ [$g \in SU(N)$]; 
on the other hand,  
the gauge symmetry $U(M)$ acts on $\Phi$ from the right: 
\begin{align}
\Phi \ &\to \ \Phi' 
\ = \ 
\Phi e^{-i \Lambda} \; , \ \ \ 
e^V \ \to \ e^{V'} \ = \ e^{i \Lambda} e^V e^{-i \Lambda^{\dagger}} \; ,
\end{align}
where $\Lambda (x, \theta, \bar{\theta})$ is a parameter 
chiral superfield, taking a value in the Lie algebra of $U(M)$. 
Note that the local invariance group is enlarged to 
the complexification of the gauge group, 
$U(M)^{\bf C}=GL(N,{\bf C})$, 
because the scalar component of 
$\Lambda (x,\theta, \bar{\theta})$ is complex. 
The Lagrangian invariant under the global $SU(N)$ and 
the local $U(M)$ symmetries is given by 
\begin{align}
{\cal L} 
\ &= \ 
\int \! d^4 \theta \,
\left[
\tr (\Phi^{\dagger} \Phi e^V) - c \, \tr V \right] \; , 
\label{compact-G}
\end{align}
where $c \, \tr V$ is the FI D-term. 

Integrating out the auxiliary vector superfield $V$,
we obtain 
\begin{align}
{\cal K} (\Phi, \Phi^{\dagger}) 
\ &= \ 
c \log \det (\Phi^{\dagger} \Phi) \; , \label{cpt-g-kahler-1}
\end{align}
where we have omitted constants, since they disappear 
under the integration over $\theta$. 
Since the gauge symmetry is complexified,
we can write the gauge as
\begin{align}
\Phi \ &= \ \begin{pmatrix}
{\bf 1}_M \cr 
 \varphi 
\end{pmatrix} \; ,
\end{align}
where $\varphi (x, \theta, \bar{\theta})$ is an $(N-M) \times M$
matrix-valued chiral superfield. 
Substituting this into (\ref{cpt-g-kahler-1}),
we obtain the \kahler potential of 
$G_{N,M}=SU(N)/[SU(N-M)\times U(M)]$:
\begin{align}
{\cal K} (\varphi, \varphi^{\dagger}) 
\ &= \ 
c \log \det \big( {\bf 1}_M + \varphi^{\dagger} \varphi \big) \; . 
\label{cpt-g-Kahler}
\end{align}

Next, we construct the non-compact Calabi-Yau manifolds
by restricting the gauge degrees of freedom from $U(M)$ to $SU(M)$.
Let $V(x,\theta,\bar{\theta}) = V^A T_A$ be a vector superfield 
taking a value in the Lie algebra of $SU(M)$, 
whose generators are $T_A$. 
The \kahler potential is
\begin{align}
 {\cal K}_0 (\Phi, \Phi^{\dagger}, V) 
 \ &= \ f (\tr (\Phi^{\dagger} \Phi e^V)) \; , 
\label{noncpt-G}
\end{align}
where $f$ is an arbitrary function\footnote{There exist independent
invariants 
${\rm tr}[(\Phi^{\dagger} \Phi e^V)^2]$, $\cdots$,
${\rm tr}[(\Phi^{\dagger} \Phi e^V)^M]$, 
besides ${\rm tr}(\Phi^{\dagger} \Phi e^V)$.
We can show that, even if these are included as the arguments of 
the arbitrary function of (\ref{noncpt-G}), 
we obtain the same result (\ref{kahler0-g}).
}.  
The equations of motion for $V$ read 
\begin{align}
\Pd{{\cal L}}{V} 
\ &= \ 
f' (\tr (\Phi^{\dagger} \Phi e^V)) 
\cdot \tr (\Phi^{\dagger} \Phi e^V T_A )
\ = \ 
0 \; , 
\end{align}
where the prime denotes the differentiation with respect to the
argument of $f$.
Then, we obtain  
\begin{align}
 f' (\tr (\Phi^{\dagger} \Phi e^V)) 
 \cdot \Phi^{\dagger} \Phi e^V 
 \ = \ C {\bf 1}_M  \; ,\label{eom-V}
\end{align}
where $C(x,\theta,\bar{\theta})$ is a vector superfield.
There is an alternative way to obtain 
this equation~\cite{HKN3}.  
The trace and the determinant of (\ref{eom-V}) are
\bsubeq
\begin{align}
f' (\tr(\Phi^{\dagger} \Phi e^V)) \cdot \tr (\Phi^{\dagger} \Phi e^V) 
\ &= \ 
M C \; , \\
\big[ f' (\tr (\Phi^{\dagger} \Phi e^V)) \big]^M 
\cdot \det (\Phi^{\dagger} \Phi) 
\ &= \ 
C^M \; ,
\end{align}
\esubeq
respectively, where $\det e^V = 1$
because of the tracelessness of the $SU(M)$ gauge field $V$.
Eliminating $C$ from these equations,
we obtain the solution of $V$ as
\begin{align}
\tr (\Phi^{\dagger} \Phi e^V) 
\ &= \ 
M \big[ \det (\Phi^{\dagger} \Phi) \big]^{\frac{1}{M}} \; .
\end{align}
Substituting this back into (\ref{noncpt-G}), 
we obtain the \kahler potential
\begin{align}
{\cal K}_0 (\Phi, \Phi^{\dagger}, V(\Phi, \Phi^{\dagger})) 
\ &= \ 
f \big( M [ \det (\Phi^{\dagger} \Phi) ]^{\frac{1}{M}} \big) 
\ \equiv \
{\cal K} (X (\Phi, \Phi^{\dagger})) \; , 
\label{kahler0-g}
\end{align}
where we have defined a vector superfield 
\begin{align}
 X (\Phi, \Phi^{\dagger}) 
 \ &\equiv \ 
 \log \det \Phi^{\dagger} \Phi \; , \label{X}
\end{align}
which is invariant under 
the global $U(N)$ and the local $SU(M)$ symmetries, 
${\cal K}(X)$ is a real function of $X$ related to $f$. 
The result (\ref{kahler0-g}) can be also obtained from the 
view point of the moduli space of 
supersymmetric gauge theories~\cite{HKLR,LT}. 
Since the gauge symmetry is complexified as 
$SU(M)^{\bf C} = SL(M, {\bf C})$,
we can write the matrix-valued chiral superfield $\Phi$ as
\begin{align}
\Phi \ &= \ \sigma \begin{pmatrix}
  {\bf 1}_M \cr
  \varphi
 \end{pmatrix} \; , \label{gf-Phi-g}
\end{align}
where
$\varphi (x, \theta , \bar{\theta})$ is an $(N-M) \times M$
matrix-valued chiral superfield whose components are written as
$\varphi_{A a}$.
Upper case $A$ and lower case $a$ run from 
$1$ to $(N-M)$ and $1$ to $M$, respectively. 
$\sigma (x, \theta, \bar{\theta})$ is a chiral superfield.
Then the invariant superfield $X$ defined in 
(\ref{X}) is decomposed as
\bsubeq \label{xi-g}
\begin{align}
X \ &= \ 
M \log |\sigma|^2 + \Psi \; , \ \ \ 
\Psi \ = \ \log \xi \; , \ \ \ 
\xi 
\ \equiv \
\det ({\bf 1}_M + \varphi^{\dagger}\varphi)  \;, 
\end{align}
\esubeq
where $\Psi$ is the \kahler potential of 
$G_{N,M}$ (\ref{cpt-g-Kahler}).
We can regard this non-compact manifold locally as 
${\bf R} \times SU(N)/[SU(N-M) \times SU(M)]$.

Before discussing the Ricci-flat condition on 
the line bundle over $G_{N,M}$,  
we consider \kahler coset spaces with other isometries 
imposing holomorphic constraints on $G_{N,M}$. 
First we prepare the Lagrangian of $G_{2N,N}$ 
(\ref{compact-G}), 
in which $\Phi( x, \theta, \bar{\theta})$ is 
a $2N \times N$ matrix-valued chiral superfield.  
We can obtain the HSS of 
$SO(2N)/U(N)$ or $Sp(N)/U(N)$ by introducing 
the superpotential  
\begin{align}
W (\Phi_0, \Phi) 
\ &= \ 
\tr \big( \Phi_0 \Phi^T J' \Phi) \; , \ \ \ 
J' \ = \ \left(
\begin{array}{cc}
{\bf 0} & {\bf 1}_N \\
\epsilon {\bf 1}_N & {\bf 0}
\end{array} \right) \; ,
\end{align}
where $\Phi_0 (x, \theta, \bar{\theta})$ is 
an $N \times N$ matrix-valued
auxiliary chiral superfield. 
Here $J'$ is the rank-$2$ invariant tensor in which 
$\epsilon$ stands for a sign: 
$\epsilon = -1$ [$\epsilon = +1$] 
corresponds to $Sp(N)/U(N)$ [$SO(2N)/U(N)$]. 
Choosing the same gauge fixing as $G_{N,M}$,
we can write $\Phi(x, \theta, \bar{\theta})$ as
\begin{align}
\Phi \ &= \ \sigma \begin{pmatrix}
  {\bf 1}_N \cr
  \varphi
 \end{pmatrix} \; . \label{gf-Phi-so-sp}
\end{align}
Here
$\varphi(x, \theta, \bar{\theta})$ is an $N \times N$ matrix-valued
chiral superfield whose components are written as $\varphi_{a b}$,
where $1 \leq a < b \leq N$ for $SO(2N)/U(N)$ 
or $1 \leq a \leq b \leq N$ for $Sp(N)/U(N)$.
These non-compact manifolds can be locally regarded as
${\bf R} \times SO(2N)/SU(N)$ and ${\bf R} \times Sp(N)/SU(N)$.

The metrics are defined in (\ref{metric})
where holomorphic coordinates are $z^{\mu} = (\sigma ,
\varphi_{A a})$ for $G_{N,M}$ and $z^{\mu} = (\sigma , \varphi_{a b})$
for $SO(2N)/U(N)$ or $Sp(N)/U(N)$.
Thus the determinants $\det g_{\mu \nu^*}$ can be written as 
\begin{align}
\det g_{\mu \nu^*} \ &= \ \left\{
\begin{array}{l@{\ls}l}
\dps \frac{M^2}{|\sigma|^2} {\cal K}'' \cdot \det_{(A a) (B b)} 
\Big(
{\cal K}' \frac{\del^2 X}{\del \varphi_{A a} \del \varphi_{B b}^*} 
\Big)
& \mbox{for $G_{N,M}$} \; , \\
\dps \frac{N^2}{|\sigma|^2} {\cal K}'' \cdot \det_{(a b) (c d)} 
\Big(
{\cal K}' \frac{\del^2 X}{\del \varphi_{a b} \del \varphi_{c d}^*} 
\Big) 
& \mbox{for $SO(2N)/U(N)$ or $Sp(N)/U(N)$} \; .
\end{array} \right. \label{det-g-so-sp0}
\end{align}
The $X$'s differentiated twice can be calculated, 
to yield
\begin{align}
\frac{\del^2 X}{\del \varphi_{A a} \del \varphi^*_{B b}}
 \ &= \ \del_{(Aa)}\del_{(Bb)^*}\Psi 
 \ = \ ({\bf 1}_M + \varphi^{\dagger} \varphi)^{-1}_{a b} 
 \Big[ {\bf 1}_{(N-M)} -
   \varphi ({\bf 1}_M + \varphi^{\dagger} \varphi)^{-1} 
   \varphi^{\dagger} \Big]_{B A} \;, \label{deriv-g}
\end{align}
for $G_{N,M}$ and
\begin{align}
\frac{\del^2 X}{\del \varphi_{a b} \del \varphi^*_{c d}} 
 \ &= \ \del_{(ab)} \del_{(cd)^*} \Psi \nonumber \\
 \ &= \ \Big( 1 - \half \delta_{a b} \Big) 
  \Big( 1 - \half \delta_{c d} \Big) 
  \Big\{ ({\bf 1}_N + \varphi^{\dagger} \varphi)^{-1}_{b d} 
   \big[ {\bf 1}_N 
     - \varphi ({\bf 1}_N + \varphi^{\dagger}\varphi)^{-1} 
       \varphi^{\dagger} \big]_{c a} \nonumber \\
 \ & \LS \ \ \ \ 
 - \epsilon \; ({\bf 1}_N + \varphi^{\dagger}\varphi)^{-1}_{b c} 
   \big[ {\bf 1}_N 
      - \varphi ({\bf 1}_N + \varphi^{\dagger} \varphi)^{-1} 
        \varphi^{\dagger} \big]_{d a}
 + (a \leftrightarrow b, c \leftrightarrow d)
 \Big\} \; , \label{deriv-so-sp}
\end{align}
for $SO(2N)/U(N)$ ($\epsilon = + 1$, $a < b$) or 
for $Sp(N)/U(N)$ ($\epsilon = - 1$, $a \leq b$). 

Under the complex isotropy transformation of $SL(N-M, {\bf C}) \times
SL(M, {\bf C})$ for $G_{N,M}$ [$SL(N, {\bf C})$ for $SO(2N)/U(N)$ or
$Sp(N)/U(N)$], 
we can calculate the determinant (\ref{det-g-so-sp0}): 
\begin{align}
\det g_{\mu \nu^*} \ &= \ \left\{
\begin{array}{l@{\ls}l}
\dps M^2 |\sigma|^{2(MN-1)} 
e^{-NX} {\cal K}'' ({\cal K}')^{M(N-M)} 
& \mbox{for $G_{N,M}$} \; , \\
\dps N^2 2^{\half N(N-1)} |\sigma|^{2N(N-1)-2} 
e^{-(N-1)X} {\cal K}'' ({\cal K}')^{\half N(N-1)} 
& \mbox{for $SO(2N)/U(N)$} \; , \\
\dps N^2 2^{\half N(N-1)} |\sigma|^{2N(N+1)-2} 
e^{-(N+1)X} {\cal K}'' ({\cal K}')^{\half N(N+1)} 
& \mbox{for $Sp(N)/U(N)$} \; , 
\end{array} \right. 
\end{align}
Therefore the Ricci-flat condition (\ref{RF-condition}) can be
solved for ${\cal K}'$ as
\begin{align}
{\cal K}' \ = \ \left\{
\begin{array}{l@{\ls}l@{\ls}l}
\dps \big( \lambda e^{N X} + b \big)^{\frac{1}{g}}
& g \ \equiv \ M(N-M) +1 & \mbox{ for $G_{N,M}$} \; ,\\
\dps \big( \lambda e^{(N-1) X} + b \big)^{\frac{1}{f}}
& f \ \equiv \ \half N (N-1) + 1 & \mbox{ for $SO(2N)/U(N)$} \; , \\
\dps \big( \lambda e^{(N+1) X} + b \big)^{\frac{1}{h}}
& h \ \equiv \ \half N (N+1) + 1 & \mbox{ for $Sp(N)/U(N)$} \; , 
\end{array} \right. \label{rf-sol-g}
\end{align}
where $\lambda$ is a constant and $b$ is an integration constant.
The explicit expression of the 
\kahler potentials can be found in \cite{HKN3}.

The Ricci-flat metric can be calculated 
by substituting the solution (\ref{rf-sol-g}) into (\ref{metric}). 
The component $g_{\sigma \sigma^*}$ can be calculated as
\begin{align}
g_{\sigma \sigma^*} \ = \ \left\{
\begin{array}{l@{\ls}l}
\dps \lambda \frac{M^2 N}{g} 
  \big( \lambda e^{NX} + b \big)^{\frac{1}{g} - 1}
  e^{N \Psi} |\sigma|^{2 MN - 2} & \mbox{ for $G_{N,M}$} \; , \\
\dps \lambda \frac{N^2(N-1)}{f} 
  \big( \lambda e^{(N-1) X} + b \big)^{\frac{1}{f} - 1} 
  e^{(N-1) \Psi} |\sigma|^{2N(N-1) - 2} & \mbox{ for
  $SO(2N)/U(N)$} \; , \\
\dps \lambda \frac{N^2(N+1)}{h} 
  \big( \lambda e^{(N+1) X} + b \big)^{\frac{1}{h} - 1} 
  e^{(N+1) \Psi} |\sigma|^{2N(N+1) - 2} & \mbox{ for
  $Sp(N)/U(N)$} \; ,  
\end{array} \right.
\end{align}
where $\Psi$ is defined in (\ref{xi-g}). 
Although this component is singular at the $\sigma = 0$ surface: 
$g_{\sigma \sigma ^*}|_{\sigma =0}=0$, 
this singularity is just a coordinate singularity of 
$z^{\mu} =(\sigma,\varphi)$.  
By performing the coordinate transformation  
\begin{align}\label{rho}
 \rho \ &\equiv \left\{
 \begin{array}{l@{\ls}l}
  \sigma^{MN}/MN & \mbox{ for $G_{N,M}$} \; , \\
  \sigma^{N(N-1)}/N(N-1) & \mbox{ for $SO(2N)/U(N)$} \; , \\
  \sigma^{N(N+1)}/N(N+1) & \mbox{ for $Sp(N)/U(N)$} \; ,
 \end{array} \right. 
\end{align}
with $\varphi$ being unchanged, 
we obtain the regular coordinates. 
Each metric in the new coordinates $z'{}^{\mu} = (\rho,\varphi)$ 
can be calculated, to give
\bsubeq
\begin{align}
g_{\rho \rho^*} \ &= \ \lambda \frac{M^2 N}{g} 
 \big( \lambda e^{N X} + b \big)^{\frac{1}{g}-1} e^{N \Psi} \; , \\
g_{\rho (B b)^*} \ &= \ \lambda \frac{M^2 N^2}{g} 
 \big( \lambda e^{N X} + b \big)^{\frac{1}{g} -1} e^{N \Psi} 
 \rho^* \cdot \del_{(B b)^*} \Psi \; , \\
g_{(A a) (B b)^*} \ &= \ \lambda \frac{M^2 N^3}{g} 
 \big( \lambda e^{N X} + b \big)^{\frac{1}{g} -1} e^{N \Psi} 
 |\rho|^2 \cdot \del_{(A a)} \Psi \del_{(B b)^*} \Psi \nonumber \\
\ & \ \ \ \ 
 + \big( \lambda e^{N X} + b \big)^{\frac{1}{g}} 
 \cdot \del_{(Aa)} \del_{(B b)^*} \Psi \; , 
\end{align}
\esubeq
for $G_{N,M}$,
\bsubeq
\begin{align}
g_{\rho \rho^*} \ &= \ \lambda \frac{N^2(N-1)}{f} 
 \big( \lambda e^{(N-1) X} + b \big)^{\frac{1}{f}-1} e^{(N-1) \Psi} \;,\\
g_{\rho (c d)^*} \ &= \ \lambda \frac{N^2 (N-1)^2}{f} 
 \big( \lambda e^{(N -1) X} + b \big)^{\frac{1}{f} -1} e^{(N-1) \Psi} 
 \rho^* \cdot \del_{(c d)^*} \Psi \; , \\
g_{(a b) (c d)^*} \ &= \ \lambda \frac{N^2 (N-1)^3}{f} 
 \big( \lambda e^{(N-1) X} + b \big)^{\frac{1}{f} -1} e^{(N-1) \Psi} 
 |\rho|^2 \cdot \del_{(a b)} \Psi \del_{(c d)^*} \Psi \nonumber \\
\ & \ \ \ \ 
 + \big( \lambda e^{(N-1) X} + b \big)^{\frac{1}{f}}
  \cdot \del_{(a b)} \del_{(c d)^*} \Psi \; , 
\end{align}
\esubeq
for $SO(2N)/U(N)$, and
\bsubeq
\begin{align}
g_{\rho \rho^*} \ &= \ \lambda \frac{N^2(N+1)}{h} 
 \big( \lambda e^{(N+1) X} + b \big)^{\frac{1}{h}-1} 
 e^{(N+1) \Psi} \; , \\
g_{\rho (c d)^*} \ &= \ \lambda \frac{N^2 (N+1)^2}{h} 
 \big( \lambda e^{(N +1) X} + b \big)^{\frac{1}{h} -1} e^{(N+1) \Psi} 
 \rho^* \cdot \del_{(c d)^*} \Psi \; , \\
g_{(a b) (c d)^*} \ &= \ \lambda \frac{N^2 (N+1)^3}{h} 
 \big( \lambda e^{(N+1) X} + b \big)^{\frac{1}{h} -1} e^{(N+1) \Psi} 
 |\rho|^2 \cdot \del_{(a b)} \Psi \del_{(c d)^*} \Psi \nonumber \\
 \ & \ \ \ \ + \big( \lambda e^{(N+1) X} + b \big)^{\frac{1}{h}}
 \cdot \del_{(a b)} \del_{(c d)^*} \Psi \; , 
\end{align}
\esubeq
for $Sp(N)/U(N)$.
They are regular in the whole region 
(including $\rho = 0$)~\cite{HKN3}.

The metrics of the submanifolds defined by 
$\rho = 0$ ($d \rho = 0$) are
\begin{align}
g_{(A a) (B b)^*}|_{\rho = 0}\,(\varphi,\varphi^*) 
 \ &= \ b^{\frac{1}{g}} \del_{(Aa)} \del_{(B b)^*} \Psi && \mbox{ for
 $G_{N,M}$} \; , \nonumber \\
g_{(a b) (c d)^*}|_{\rho = 0}\,(\varphi,\varphi^*)  
 \ &= \ b^{\frac{1}{f}} \del_{(a b)} \del_{(c d)^*} \Psi && \mbox{ for
 $SO(2N)/U(N)$} \; , \\
g_{(a b) (c d)^*}|_{\rho = 0}\,(\varphi,\varphi^*)  
 \ &= \ b^{\frac{1}{h}} \del_{(a b)} \del_{(c d)^*}\Psi && \mbox{ for
 $Sp(N)/U(N)$} \; . \nonumber
\end{align}
Since $\Psi$ is the \kahler potential of 
$G_{N,M}$, $SO(2N)/U(N)$ or $Sp(N)/U(N)$,  
this is its metric given in  
(\ref{deriv-g}) or (\ref{deriv-so-sp}). 
Therefore we find that 
the total space is the complex line bundle 
over $G_{N,M}$, $SO(2N)/U(N)$ or $Sp(N)/U(N)$ as 
a base manifold
with the fiber $\rho$.
In the limit of $b \to 0$, each base manifold shrinks to zero-size  
and the conical singularity due to the identification 
(\ref{rho}) appears. 
If $b\neq 0$ the conical singularity is resolved by
$G_{N,M}$, $SO(2N)/U(N)$ or $Sp(N)/U(N)$ of a radius $b^{1/2g}$,
$b^{1/2f}$ or $b^{1/2h}$, respectively. 


\section{Conclusion} \label{conclusion}

In this paper 
we have constructed non-compact Calabi-Yau manifolds
interpreted as the complex line bundles over HSS.
We have presented the Ricci-flat metrics and their \kahler
potentials on these manifolds.
In particular, we have presented the new Ricci-flat metrics on
the non-compact Calabi-Yau manifolds with the isometries 
of the exceptional groups $E_6$ and $E_7$. 

There are several essential points for obtaining these potentials.
First, 
the $U(1)_{\rm local}$ symmetry, which was gauged for
obtaining the compact \kahler manifolds,
has been treated as a global symmetry 
to obtain the non-compact manifolds. 
Second,
we have performed the complex isotropy transformations 
in order to calculate the determinants of the metrics.
Using these transformations, 
we obtain the ordinary differential
equations for solving the Ricci-flat condition.
The form of the solutions is written as 
\begin{align}
{\cal K}' \ &= \ \big( \lambda e^{{\cal C} X} + b \big)^{\frac{1}{D}}
\; , \label{rfk}
\end{align}
Here $D$ is the complex dimension of the complex line bundle;
${\cal C}$ is a constant related to $N$ (see, Table \ref{HSS-table});
$\lambda$ is a constant and 
$b$ is an integration constant regarded as the resolution parameter
of the conical singularity.
Third,
we have transformed the fiber coordinates from $\sigma$ to $\rho$
to eliminate the coordinate singularity.
The metrics in new coordinates $(\rho, \varphi)$ are regular. 
On the basis of these three significant points,
we have obtained the complex line bundles over HSS, 
as summarized in Table \ref{HSS-table}.

\begin{table}[h]
\begin{center}
\begin{tabular}{c|c|c|c|c} 
type & ${\bf C} \ltimes G/H$ & $D$ & ${\cal C}$
& coordinate transformation \\ \hline\hline 
${\rm AIII}_1$ & ${\bf C} \ltimes {\bf C}P^{N-1}$ & $1 + (N-1)$ 
& $N$
& $\rho \sim \sigma^N$ \\
${\rm AIII}_2$ & ${\bf C} \ltimes G_{N,M}$ & $1 + M(N-M)$ 
& $N$
& $\rho \sim \sigma^{MN}$ \\
${\rm BDI}$ & ${\bf C} \ltimes Q^{N-2}$ & $1 + (N-2)$ 
& $N-2$
& $\rho \sim \sigma^{N-2}$ \\
${\rm CI}$ & ${\bf C} \ltimes Sp(N) / U(N)$ 
& $1 + \half N(N+1)$ 
& $N+1$
& $\rho \sim \sigma^{N(N+1)}$ \\
${\rm DIII}$ & ${\bf C} \ltimes SO(2N) / U(N)$ 
& $1 + \half N(N-1)$ 
& $N-1$
& $\rho \sim \sigma^{N(N-1)}$ \\
${\rm EIII}$ & ${\bf C} \ltimes E_6 / [SO(10) \times U(1)]$ 
& $1 + 16$ 
& $12$
& $\rho \sim \sigma^{12}$ \\
${\rm EVII}$ & ${\bf C} \ltimes E_7 / [E_6 \times U(1)]$ 
& $1 + 27$ 
& $18$
& $\rho \sim \sigma^{18}$ \\ \hline
\end{tabular}
\caption{Line bundles over Hermitian symmetric spaces.
We write the classification of coset spaces by Cartan, 
complex dimensions and the coordinate transformation to 
avoid the coordinate singularity. 
The notation $X \ltimes Y$ is used to signify a bundle over $Y$ with
fiber $X$.
The numbers $D$ and ${\cal C}$ are defined in (\ref{rfk}).}

\label{HSS-table}
\end{center}
\end{table}

Before closing the conclusion,
let us make a comment.
We have sets of the isomorphism between the lower dimensional
base manifolds~\cite{HKN3}:
\bsubeq
\begin{align}
({\rm i}) \ \ &{\bf C}P^1 \ \simeq \ SO(4)/U(2) \ \simeq \ Sp(1) /
U(1) \ \simeq \ Q^1 \; , \\
({\rm ii}) \ \ &{\bf C}P^3 \ \simeq \ SO(6) / U(3) \; , \\
({\rm iii}) \ \ &Sp(2)/U(2) \ \simeq \ Q^3 \; , \\
({\rm iv}) \ \ &G_{4,2} \ \simeq \ Q^4 \; ,
\end{align}
in addition to the novel duality relation
\begin{align}
({\rm v}) \ \ &G_{N,M} \ \simeq \ G_{N,N-M} \; .
\end{align}
\esubeq
We would like to note that each isomorphism of a pair 
of base manifolds consistently leads the isomorphism of 
the line bundles over these base manifolds~\cite{HKN3}.

\section*{Acknowledgements}

We would like to thank Takashi Yokono for helpful comments.
This work was supported in part by the Grant-in-Aid for Scientific
Research.
The work of M.N. was supported by the U.S. 
Department of Energy under grant DE-FG02-91ER40681 (Task B). 


\begin{appendix}

\section{$SO(10)$ $\gamma$-matrices} \label{SO10}

In this appendix we introduce the $SO(10)$ $\gamma$-matrices and the
charge conjugation matrices in the Weyl spinor basis.
Further discussions are given in \cite{KS}.

We define the $SO(10)$ $\gamma$-matrices $\sigma_A$ 
in the Weyl spinor basis 
($A = (m, \mu)$, $m=1,2,\cdots, 6$ and $\mu = 7,8,9,10$):
\bsubeq
\label{pauli}
\begin{align}
\sigma_A \ &= \ \big( \sigma_m , \sigma_{\mu} \big) \; , \ \
\sigma_m \ = \ \left(
\begin{array}{cc}
0 & {}^6 \sigma_m \otimes {\bf 1}_2 \\
- {}^6 \sigma_m^{\dagger} \otimes {\bf 1}_2 & 0 
\end{array} \right) \; , \ \ \ \sigma_{\mu} \ = \ \left(
\begin{array}{cc}
{\bf 1}_4 \otimes {}^4 \sigma_{\mu} & 0 \\
0 & {\bf 1}_4 \otimes {}^4 \sigma_{\mu}^{\dagger}
\end{array} \right) \; ,
\end{align}
where ${}^6 \sigma_m$ are $SO(6)$ $\gamma$-matrices in the Weyl spinor
basis 
and ${}^4 \sigma_{\mu}$ are $SO(4)$ $\gamma$-matrices in the Weyl
spinor basis, say, $2 \times 2$ Pauli matrices.
They are defined as follows:
\begin{align}
{}^6 \sigma_m 
\ &= \ 
\big( {}^6 \sigma_i , {}^6 \sigma_{i+3} \big) \; , \ \ \ 
i \ = \ 1,2,3 \; , \\
({}^6 \sigma_i)_{\alpha \beta} 
\ &= \ 
\ve_{i 4 \alpha \beta} + \delta_{\alpha}^i \delta_{\beta}^4 
- \delta_{\alpha}^4 \delta_{\beta}^i \; , \ \ \ 
({}^6 \sigma_{i+3}) 
\ = \ 
i \ve_{i 4 \alpha \beta} 
- i \delta_{\alpha}^i \delta_{\beta}^4 
+ i \delta_{\alpha}^4 \delta_{\beta}^i \; , \\
{}^4 \sigma_{\mu} 
\ &= \ 
\big( - i \sigma_1 , -i \sigma_2 , -i \sigma_3 , {\bf 1}_2 \big) \; ,
\end{align}
\esubeq
where $\ve_{ijkl}$ is the rank-4 totally anti-symmetric tensor
($\ve_{1234} = 1$).
Using these $\gamma$-matrices, we obtain the various matrices.
The spinor rotation matrices $\sigma_{AB}$ are defined by
\begin{align}
\sigma_{AB} 
\ &= \ 
\frac{1}{2} \big( 
\sigma_A \sigma_B^{\dagger} - \sigma_B \sigma_A^{\dagger} 
\big) \; .
\end{align}

The $SO(10)$ charge conjugation matrix in the Weyl spinor basis is
defined by 
\begin{align}
C \ &= \ \left(
\begin{array}{cc}
0 & - {\bf 1}_4 \otimes i \sigma_2 \\
{\bf 1}_4 \otimes i \sigma_2 & 0 
\end{array} \right) \ = \ C^T \ = \ C^{-1} \ = \ C^{\dagger} \;
. \label{cc-matrix}
\end{align}
Because of (\ref{pauli}) and (\ref{cc-matrix}), 
the combination of following matrices are symmetric:
\begin{align}
(C \sigma_A^{\dagger})^{\alpha \beta} 
\ &= \ 
(C \sigma_A^{\dagger})^{\beta \alpha} \; , \ \ \ 
(\sigma^A C^{\dagger})^{\alpha \beta} 
\ = \ 
(\sigma^A C^{\dagger})^{\beta \alpha} \; .
\end{align}


\section{$E_6$ and $E_7$ Algebras} \label{e-algebra}

In this appendix, 
we show the short review of the $E_6$ and $E_7$ algebras 
constructed by their maximal subgroups.
More discussions are expanded in \cite{IKK2,KS,HN1}.


\subsection{Construction of $E_6$ algebra}

Since the decomposition of the adjoint representation of $E_6$ under its
maximal subgroup $SO(10) \times U(1)$ is 
${\bf 78} = ({\bf 45},0) \oplus ({\bf 1},0)
\oplus({\bf 16},1)\oplus ({\bf \ol{16}},-1)$, 
we construct the $E_6$ algebra as 
${\cal E}_6 = {\cal SO}(10) \oplus {\cal U}(1) 
\oplus {\bf 16} \oplus {\bf \ol{16}}$: 
We prepare the $SO(10)$ generators $T_{AB}$, 
the $U(1)$ generator $T$, 
spinor generators $E_{\alpha}$ and 
$\ol{E}^{\alpha} = (E_{\alpha})^{\dagger}$,
belonging to ${\bf 16}$ and $\ol{\bf 16}$, respectively.
Then their commutation relations can be calculated 
as follows: 
\begin{align}
&[T_{AB} , T_{CD}] 
 \ = \ -i (\delta_{BC} T_{AD} + \delta_{AD} T_{BC}
    - \delta_{AC} T_{BD} - \delta_{BD} T_{AC}) , 
   \ \ \ [T, T_{AB}] \ = \ 0 \; , \nonumber \\
&[T_{AB} , E_{\alpha}] 
  \ = \ - {({\sigma_{AB}})_{\alpha}}^{\beta} E_{\beta} \; , 
\ \ \ [T_{AB},\ol{E}^{\alpha}] 
\ = \ {({\sigma^*_{AB}})^{\alpha}}_{\beta} \ol{E}^{\beta} \; ,
 \nonumber \\
&[T, E_{\alpha}] \ = \ \frac{\sqrt{3}}{2} E_{\alpha}, 
\ \ \  [T, \ol{E}_{\alpha}] \ = \ - \frac{\sqrt{3}}{2} \ol{E}^{\alpha}
 \; , \nonumber \\
&[E_{\alpha},E_{\beta}] 
  \ = \ [\ol{E}^{\alpha},\ol{E}^{\beta} ] \ = \ 0 \; , 
\ \ \  [E_{\alpha} ,\ol{E}^{\beta} ]
 \ = \ -\half {({\sigma_{AB}})_{\alpha}}^{\beta} T_{AB} 
   + \frac{\sqrt{3}}{2} {\delta_{\alpha}}^{\beta} T \; .
\end{align}
The $U(1)$ charge of $E_{\alpha}$ is determined by 
the difference between $U(1)$ charges 
of $x$ and $y$ or $y$ and $z$ in (\ref{E6-tr.}): 
$\frac{2}{\sqrt{3}} - \frac{1}{2\sqrt{3}} 
= \frac{1}{2\sqrt{3}} - (-\frac{1}{\sqrt{3}})= \frac{\sqrt{3}}{2}$.
The second coefficient of the last equation has 
the same value as the $U(1)$ charge of $E_{\alpha}$, from 
the anti-symmetric property of the structure constants. 
The relative weight of the first and the second terms  
is determined by the Jacobi identity, 
$[\ol{E},[E,E]] + {\rm (cyclic)} = 0$, 
and the non-trivial identity for 
the spinor generators:
$\sum_{A,B}({\sigma_{AB}})_{\alpha}{}^{[\beta}
 ({\sigma_{AB}})_{\gamma}{}^{\delta]}
  \ = \ \frac{3}{2} \delta_{\alpha}{}^{[\beta}
 \delta_{\gamma}{}^{\delta]}$.

The transformation law of $\vec{\phi}$ under 
the complex extension of $E_6$ is
\begin{align}
\delta \vec{\phi}
\ &= \ \left(i \theta T + \frac{i}{2} \theta_{AB} T_{AB} 
    + \ol{\epsilon}^{\alpha} E_{\alpha} 
    + \epsilon_{\alpha} \ol{E}^{\alpha} \right) 
  \vec{\phi} \nonumber \\
\ &= \ \left(
\begin{array}{ccc} 
\frac{2 i}{\sqrt{3}} \theta & \ol{\epsilon}^{\beta} & {\bf 0} \\
\epsilon_{\alpha} & 
\frac{i}{2} \theta_{AB}{(\sigma_{AB})_{\alpha}}^{\beta} 
+ \frac{i}{2\sqrt{3}} \theta \delta_{\alpha}^{\beta} & 
-\frac{1}{\sqrt{2}} (\ol{\epsilon} \sigma_B C)_{\alpha} \\
{\bf 0} & - \frac{1}{\sqrt{2}} (C \sigma_A^{\dagger} \epsilon )^{\beta} &
          \theta_{AB} - \frac{i}{\sqrt{3}} \theta \delta_{AB}
\end{array} \right) \left(
\begin{array}{c}
x \\
y_{\beta} \\
z^B 
\end{array} \right) \; , \label{E6-tr.}
\end{align}
where $\frac{i}{2} \theta_{CD} \rho{(T_{CD})^A}_B = \theta_{AB}$, 
and $\rho(T_{AB})$ are the vector representation 
matrices of $SO(10)$. 
The $16 \times 16$ matrices $\sigma_A$, $\sigma_{AB}$ and $C$ are 
$SO(10)$ $\gamma$-matrices, 
spinor rotation matrices 
and the charge conjugation matrix, respectively (defined in appendix
\ref{SO10}). 
Normalizations are fixed by 
$\tr T^2 = \tr (T_{AB})^2 
= \tr E_{\alpha} \ol{E}^{\alpha}=6$ (no summation).
In (\ref{E6-tr.}) 
$\ol{\epsilon}_{\alpha}$ are independent of $\epsilon^{\alpha}$ 
if we consider the action of $E_6{}^{\bf C}$, 
while $\ol{\epsilon}_{\alpha} = \epsilon^{*{\alpha}}$ hold 
when we consider the real group $E_6$.


\subsection{Construction of $E_7$ algebra}

The decomposition of the adjoint representation of $E_7$ 
under the maximal subgroup $E_6 \times U(1)$ is 
${\bf 133} = ({\bf 78},0) \oplus ({\bf 1},0)
\oplus ({\bf 27},1) \oplus ({\bf \ol{27}},-1)$, 
where the second components are the $U(1)$ charges.
Hence, we can construct the $E_7$ algebra 
by adding generators $E^i$ and $\ol{E}_i (= (E^i)^{\dagger})$ 
($i=1,\cdots,27$), which belong to 
the $E_6$ fundamental and anti-fundamental 
representations, respectively, 
to the $E_6 \times U(1)$ algebra, 
$T_A$ ($A=1,\cdots,78$) and $T$: 
${\cal E}_7 = {\cal E}_6 \oplus {\cal U}(1) 
\oplus {\bf 27} \oplus {\bf \ol{27}}$.
In the same manner as we constructed 
the $E_6$ algebra, 
their commutation relations are obtained as follows: 
\begin{align}
&[T_A,T_B] \ = \ i {f_{AB}}^C T_C \; , && [T, T_A] \ = \ 0 \; ,
\nonumber \\
&[T_A,E^i] \ = \ {\rho(T_A)^i}_j E^j \; , &&
   [T_A,\ol{E}_i] \ = \ -{{\rho(T_A)^T}_i}^j \ol{E}_j \; , \nonumber \\
&[T, E^i] \ = \ \sqrt{\frac{2}{3}} E^i \; , &&
   [T, \ol{E}_i] \ = \ - \sqrt{\frac{2}{3}} \ol{E}_i \; , \nonumber \\
&[E^i,E^j] \ = \ [\ol{E}_i,\ol{E}_j ] \ = \ 0 \; , &&
[E^i,\ol{E}_j] \ = \ {\rho(T_A)^i}_j T_A + \sqrt{\frac{2}{3}}
{\delta^i}_j T \; .
\end{align}
Here $\rho(T_A)$ are fundamental representation matrices, 
and the $f_{AB}{}^C$ are structure constants of $E_6$.
The $U(1)$ charge of $E^i$ is determined from 
the difference of $x$ and $y^i$, etc., in (\ref{E7_action}), 
and the $U(1)$ charge of $\ol{E}_i$ is its conjugate.  
In the last equation, 
the coefficient of the second term coincides with 
the $U(1)$ charge of $E^i$ due to the anti-symmetricity 
of the structure constants of $E_7$. 
The first term is determined by 
the Jacobi identity $[\ol{E},[E,E]] + ({\rm cyclic}) = 0$ and 
the non-trivial identity for 
the $E_6$ fundamental representation,
$\sum_A \; {\rho(T_A)^{[i}}_j {\rho(T_A)^{k]}}_l
  \ = \ -\frac{2}{3} {\delta^{[i}}_j {\delta^{k]}}_l$.

The action of the $E_7$ algebra on 
the fundamental representation is 
\begin{align}
\delta \vec{\phi}
\ &= \ \Big( i \theta T + i \theta_A T_A 
    + \ol{\epsilon}_i E^i + \epsilon^i \ol{E}_i \Big) \vec{\phi}
\nonumber \\
\ &= \ \left(
\begin{array}{cccc}
i \sqrt{\frac{3}{2}} \theta & \ol{\epsilon}_j  & {\bf 0} & 0 \\
\epsilon^i & i \theta_A {\rho(T_A)^i}_j + i \sqrt{\frac{1}{6}} \theta
{\delta^i}_j & {\Gamma}^{ijk} \ol{\epsilon}_k & {\bf 0} \\
  {\bf 0} & \Gamma_{ijk} \epsilon^k & - i \theta_A {{\rho({T_A})^T}_i}^j 
- i\sqrt{\frac{1}{6}} \theta {\delta_i}^j & \ol{\epsilon}_i \\
   0 & {\bf 0} & \epsilon^j & -i \sqrt{\frac{3}{2}}\theta 
\end{array} \right) \left(
\begin{array}{c}
x \\
y^j \\
z_j \\
 w
\end{array} \right) \; , \label{E7_action}
\end{align}
where $\rho(T_A)$ are the $27 \times 27$ representation matrices
for the fundamental representation, 
$\Gamma_{ijk}$ is the $E_6$ invariant tensor 
and $\Gamma^{ijk}$ is its conjugate.
Here normalizations have been determined by 
$\tr T^2 = \tr (T_A)^2 = \tr E^i \ol{E}_i = 12$ 
(no summation). 
In (\ref{E7_action}) 
$\ol{\epsilon}_i$ are independent of 
$\epsilon^i$ 
if we consider the action of $E_7{}^{\bf C}$, 
while $\ol{\epsilon}_i = \epsilon^i$ hold 
when we consider the real group $E_7$. 

\end{appendix}


\end{document}